\documentclass[11pt]{article}
\pdfoutput=1
\usepackage[T1]{fontenc}
\usepackage{jheppub}
\usepackage[ascii]{inputenc}
\usepackage{xcolor,tikz,graphicx,dcolumn,bm,bbm}
\usepackage[capitalize]{cleveref}
\usepackage[USenglish]{babel}
\usepackage[T1]{fontenc}
\usepackage{microtype}

\newcommand{\ot}{\otimes}
\newcommand{\eps}{\varepsilon}
\newcommand{\id}{\mathbbm 1}
\newcommand{\numdomains}{r}
\DeclareMathOperator{\tr}{tr}
\DeclareMathOperator{\NC}{NC}

\DeclareMathOperator{\rk}{rank}
\DeclareMathOperator{\Var}{Var}

\usepackage{braket}
\allowdisplaybreaks[4]
\def\be{\begin{equation}}
\def\ee{\end{equation}}
\def\ba#1\ea{\begin{align}#1\end{align}}
\def\bg#1\eg{\begin{gather}#1\end{gather}}
\def\bm#1\em{\begin{multline}#1\end{multline}}
\def\bmd#1\emd{\begin{multlined}#1\end{multlined}}

\def\g{\gamma}

\def\r{\rho}

\def\S{\Sigma}
\def\t{\tau}

\def\x{\xi}

\def\y{\psi}

\def\la{\label}

\def\er{\eqref}

\def\fr{\frac}

\def\pa{\partial}

\def\ol{\overline}

\def\eq{\equiv}

\def\cd{\cdots}
\def\ap{\approx}

\def\({\left(}
\def\){\right)}
\def\[{\left[}
\def\]{\right]}
\def\<{\langle}
\def\>{\rangle}

\def\bZ{{\mathbb Z}}
\def\cB{{\mathcal B}}
\def\cH{{\mathcal H}}
\newcommand{\ext}{{\text{ext}}}
\newcommand{\even}{{(\text{even})}}
\newcommand{\odd}{{(\text{odd})}}
\newcommand{\sym}{{(\text{sym})}}
\newcommand{\nsym}{{(\text{nsym})}}

\begin{document}

\title{Holographic entanglement negativity and replica symmetry breaking}
\author[a]{Xi Dong,}
\affiliation[a]{Department of Physics, University of California, Santa Barbara, CA 93106, USA}
\emailAdd{xidong@ucsb.edu}
\author[b]{Xiao-Liang Qi,}
\affiliation[b]{Stanford Institute for Theoretical Physics, Physics Department, Stanford University, Stanford, CA 94305, USA}
\emailAdd{xlqi@stanford.edu}
\author[c]{and Michael Walter}
\emailAdd{m.walter@uva.nl}
\affiliation[c]{Korteweg-de Vries Institute for Mathematics, Institute for Theoretical Physics, Institute for Logic, Language, and Computation, QuSoft, University of Amsterdam, The Netherlands}
\abstract{
Since the work of Ryu and Takayanagi, deep connections between quantum entanglement and spacetime geometry have been revealed.
The negative eigenvalues of the partial transpose of a bipartite density operator is a useful diagnostic of entanglement.
In this paper, we discuss the properties of the associated \emph{entanglement negativity} and its R\'enyi generalizations in holographic duality.
We first review the definition of the R\'enyi negativities, which contain the familiar logarithmic negativity as a special case.
We then study these quantities in the random tensor network model and rigorously derive their large bond dimension asymptotics.
Finally, we study entanglement negativity in holographic theories with a gravity dual, where we find that R\'enyi negativities are often dominated by bulk solutions that break the replica symmetry.
From these replica symmetry breaking solutions, we derive general expressions for R\'enyi negativities and their special limits including the logarithmic negativity.
In fixed-area states, these general expressions simplify dramatically and agree precisely with our results in the random tensor network model.
This provides a concrete setting for further studying the implications of replica symmetry breaking in holography.}
\maketitle

\section{Introduction}\label{sec:intro}
Holographic duality~\cite{maldacena1999large,witten1998anti,gubser1998gauge} is a duality between $d$-di\-men\-sion\-al field theory and $(d+1)$-dimensional gravitational theories in asymptotically anti-de Sitter (AdS) space.
In~2006, the discovery of the Ryu-Takayanagi (RT) formula~\cite{ryu2006holographic} introduced quantum entanglement as a key ingredient in the holographic dictionary.
The RT formula and its generalizations~\cite{hubeny2007covariant,dong2014holographic} relate the entanglement entropy of a boundary region to the area of the extremal surface in the bulk that is homologous to the same region.
A natural question is whether other quantum information measures also have geometrical counterparts in the holographic dual theory.
Various quantities have been studied in the literature such as R\'enyi entropies~\cite{Lewkowycz:2013nqa,dong2016gravity}, relative entropy~\cite{blanco2013relative,jafferis2016relative}, the entanglement of purification~\cite{takayanagi2017holographic,nguyen2018entanglement}, and the reflected entropy~\cite{dutta2019canonical,Kusuki:2019rbk,jeong2019reflected}, just to name a few.

In this paper, we study entanglement negativity and its R\'enyi generalization in the holographic duality and its random tensor network toy model.
The negativity is a measure of quantum entanglement in mixed states.
We begin by reviewing its definition.
Given a density operator~$\rho_{AB}$ with a bipartition into $A$ and $B$, we choose an orthonormal basis~$\ket a$ for~$A$ and an orthonormal basis~$\ket b$ for~$B$.
Define the partial transpose~$\rho_{AB}^{T_B}$ as the operator obtained by taking a transpose on the $B$-system, as follows:%
\footnote{The definition of $\rho^{T_B}_{AB}$ depends on a choice of basis, but the eigenvalues of $\rho^{T_B}_{AB}$ are invariant under basis change, and are thus intrinsic properties of $\rho_{AB}$.}
\begin{align*}
  \braket{a,b|\rho^{T_B}_{AB}|a',b'} \equiv \braket{a,b'|\rho_{AB}|a',b}.
\end{align*}
The resulting operator $\rho_{AB}^{T_B}$ is still Hermitian with trace one, so has real eigenvalues $\{\lambda_i\}_{i=1}^{D_{AB}}$ summing to one.
While the eigenvalues of a density matrix are all non-negative, this is not necessarily true for the partial transpose.
For example, if $\rho_{AB}$ is an EPR pair of two qubits, the eigenvalues of~$\rho_{AB}^{T_B}$ are $\{\frac12,\frac12,\frac12,-\frac12\}$.
If $\rho_{AB}$ is unentangled (separable), however, it is easy to see that $\rho_{AB}^{T_B}$ remains a positive semidefinite operator.
Thus, negative eigenvalues in the partial transpose serve as a diagnostic of entanglement~\cite{peres1996separability}.
This motivates the \emph{negativity}~\cite{vidal2002computable} and the \emph{logarithmic negativity}~\cite{plenio2005logarithmic}, which are defined by
\begin{align}\label{eq:N and E_N}
  N(\rho_{AB}) \equiv \sum_{i=1}^{D_{AB}} \frac{|\lambda_i| - \lambda_i}2 = \sum_{i : \lambda_i < 0} |\lambda_i|, \quad
  E_N(\rho_{AB}) \equiv \log \sum_{i=1}^{D_{AB}} |\lambda_i| = \log\left( 2N(\rho_{AB}) + 1 \right),
\end{align}
respectively.%
\footnote{The logarithmic negativity is often defined with logarithm to base two. Here we use the natural logarithm.}
Both quantities are entanglement monotones, and the logarithmic negativity is an upper bound on the distillable entanglement~\cite{vidal2002computable,plenio2005logarithmic,audenaert2003entanglement}.
If either quantity is positive then $\rho_{AB}$ is necessarily entangled.%
\footnote{There are, however, entangled states with positive semidefinite partial transpose. Such PPT states are bound entangled, which means that no entanglement can be distilled from them.}
The logarithmic negativity and related negativity measures have been discussed in a number of interesting prior works~\cite{Calabrese:2012ew,Calabrese:2012nk,Rangamani:2014ywa,Calabrese:2014yza,Chaturvedi:2016rft,Chaturvedi:2016rcn,Jain:2017aqk,Jain:2017xsu,Malvimat:2017yaj,Kudler-Flam:2018qjo,Tamaoka:2018ned,Malvimat:2018txq,Kudler-Flam:2019wtv,Kusuki:2019zsp,Basak:2020bot,Basak:2020oaf,Kudler-Flam:2020xqu,Lu:2020jza,Basak:2020aaa}.

We begin in \cref{sec:renyi} with a review of the R\'enyi generalizations of negativity measures.
We then study these quantities in the random tensor network model and rigorously derive their large bond dimension asymptotics in \cref{sec:rtn}.
We study entanglement negativity in holographic theories with a gravity dual in \cref{sec:holo}.
We close in \cref{sec:discussion} with a summary and discussion of the relation between results in prior works and our findings.

\section{R\'enyi negativities}\label{sec:renyi}
Just like one can generalize the von Neumann entropy to the R\'enyi entropies, $S^{(k)}(\rho)=\frac1{1-k}\log\tr[\rho^k]$, one can also define R\'enyi generalizations of negativity measures:
\begin{align}
    N_k(\rho_{AB})={\rm tr}\left( \bigl(  \rho_{AB}^{T_B} \bigr)^k\right)\label{eq:Nk}
\end{align}
We will call~$N_k$ the \emph{$k$-th R\'enyi negativity}.
Since the eigenvalues of $\rho_{AB}^{T_B}$ can be negative, the analytic continuation of $N_k$ to real~$k$ needs to be done separately for even and odd~$k$:
\begin{align}
    N_{2n}^\even (\rho_{AB}) &=\sum_i\left|\lambda_i\right|^{2n},\la{neven}\\
    N_{2n-1}^\odd (\rho_{AB}) &=\sum_i{\rm sgn}\left(\lambda_i\right)\left|\lambda_i\right|^{2n-1}\la{nodd}.
\end{align}
Then the \emph{logarithmic negativity} is the $k\to1$ ($n\to\frac12$) limit of the logarithm of the analytic continuation for even~$k$, since it only depends on the absolute values of the eigenvalues:
\begin{align}\la{en}
    E_N\left(\rho_{AB}\right)=\lim_{n\rightarrow \frac12}\log N_{2n}^\even \left(\rho_{AB}\right).
\end{align}
The \emph{negativity} can be directly extracted from this since $E_N(\rho_{AB}) = \log ( 2 N(\rho_{AB}) + 1 )$; cf.~\cref{eq:N and E_N}.
The value of $\log N_{2n-1}^\odd$ for $n\rightarrow 1$ vanishes, since ${\rm tr}\rho^{T_B}_{AB}={\rm tr}\rho_{AB}=1$.
It is natural to consider its derivative with $n$, as an analog of the von Neumann entropy:
\be\label{st}
S^{T_B}_{AB}
= -\lim_{n\to1} \frac12 \pa_n \log N_{2n-1}^\odd
= -\sum_i\lambda_i\log\left|\lambda_i\right|.
\ee
We call this quantity the \textit{partially transposed entropy}.%
\footnote{This quantity was studied in~\cite{Tamaoka:2018ned} as well and was called the ``odd entanglement entropy'' there.
Here we call it the ``partially transposed entropy'' to emphasize that it is an analogue of the entropy for the partially transposed density operator.}
More generally, we define the \emph{refined R\'enyi negativities} in the even and odd case by
\begin{align}
\la{ste}
S^{T_B(k,\text{even})}_{AB} &= -k^2 \pa_k \(\fr1k \log N_k^\even\),\\
\la{sto}
S^{T_B(k,\text{odd})}_{AB} &= -k^2 \pa_k \(\fr1k \log N_k^\odd\),
\end{align}
in analogy to the refined R\'enyi entropies of~\cite{dong2016gravity}.
The partially transposed entropy \er{st} is simply the $k\to 1$ limit of the refined odd R\'enyi negativity~\eqref{sto}.
Another interesting quantity is the $k\to2$ limit of the refined even R\'enyi negativity~\eqref{ste}, which can also be computed as follows,
\begin{align}\label{st2}
  S^{T_B(2)}_{AB}
= -\lim_{n\to1} n^2 \pa_n \(\fr{1}{n} \log N_{2n}^\even\)
= -\sum_i\frac{\lambda_i^2}{\sum_j\lambda_j^2}\log\frac{\lambda_i^2}{\sum_j\lambda_j^2},
\end{align}
and hence agrees with the von Neumann entropy of the density matrix $(\r_{AB}^{T_B})^2$ once it is properly normalized.

In the rest of the paper, we will investigate $N_k$ and its different limits introduced above.
It is helpful to introduce an alternative expression of $N_k$ for integer $k$ as the expectation value of a particular permutation operator in $k$ copies of the original system:
\begin{align}
    N_k={\rm tr}\left[\rho_{AB}^{\otimes k} \left( P_A(X) \otimes P_B(X^{-1}) \right) \right]\label{eq:Nk2}.
\end{align}
Here, $\rho_{AB}^{\otimes k}$ is the tensor product of $k$~copies of the original density operator, $X$ is a $k$-cycle (for definiteness, we take the standard one) and $X^{-1}$ its inverse, and $P_A(X)$ and $P_B(X^{-1})$ are both special cases of the general notation~$P_M(g)$, by which we denote the representation of a permutation group element~$g\in S_k$ on the~$k$ copies of some subsystem~$M$.

\section{Negativity in random tensor networks}\label{sec:rtn}
Tensor network states are quantum many-body states constructed from few-body entangled states~\cite{murg2009tnsreview,orus2014tnsreview}.
With their entanglement properties constrained by the network geometry, tensor networks become natural toy models for relating entanglement with geometry~\cite{swingle2012entanglement,vidal2007entanglement,vidal2008class,qi2013exact,pastawski2015holographic,hayden2016holographic,nezami2016multipartite,yang2016bidirectional}.
Ref.~\cite{hayden2016holographic} proposed the \emph{random tensor network} (RTN) model, which provides a toy model of the holographic duality that reproduces many of its qualitative features.
An RTN is defined by assigning random-valued tensors on the nodes of a given (non-random) graph.
Contracting the indices on each link leads to a quantum state defined on the boundary (i.e., the dangling legs of the graph).
In the limit of large bond dimension, RTNs satisfies the RT formula with quantum corrections, and the operator mapping between bulk and boundary satisfies the same local reconstruction~\cite{dong2016reconstruction} and quantum error correction properties~\cite{almheiri2015bulk} as the holographic duality (to leading order in the large~$N$ limit).

\subsection{R\'enyi negativities in random tensor networks}

\begin{figure}[t]
\centering
\includegraphics[width=3.3in]{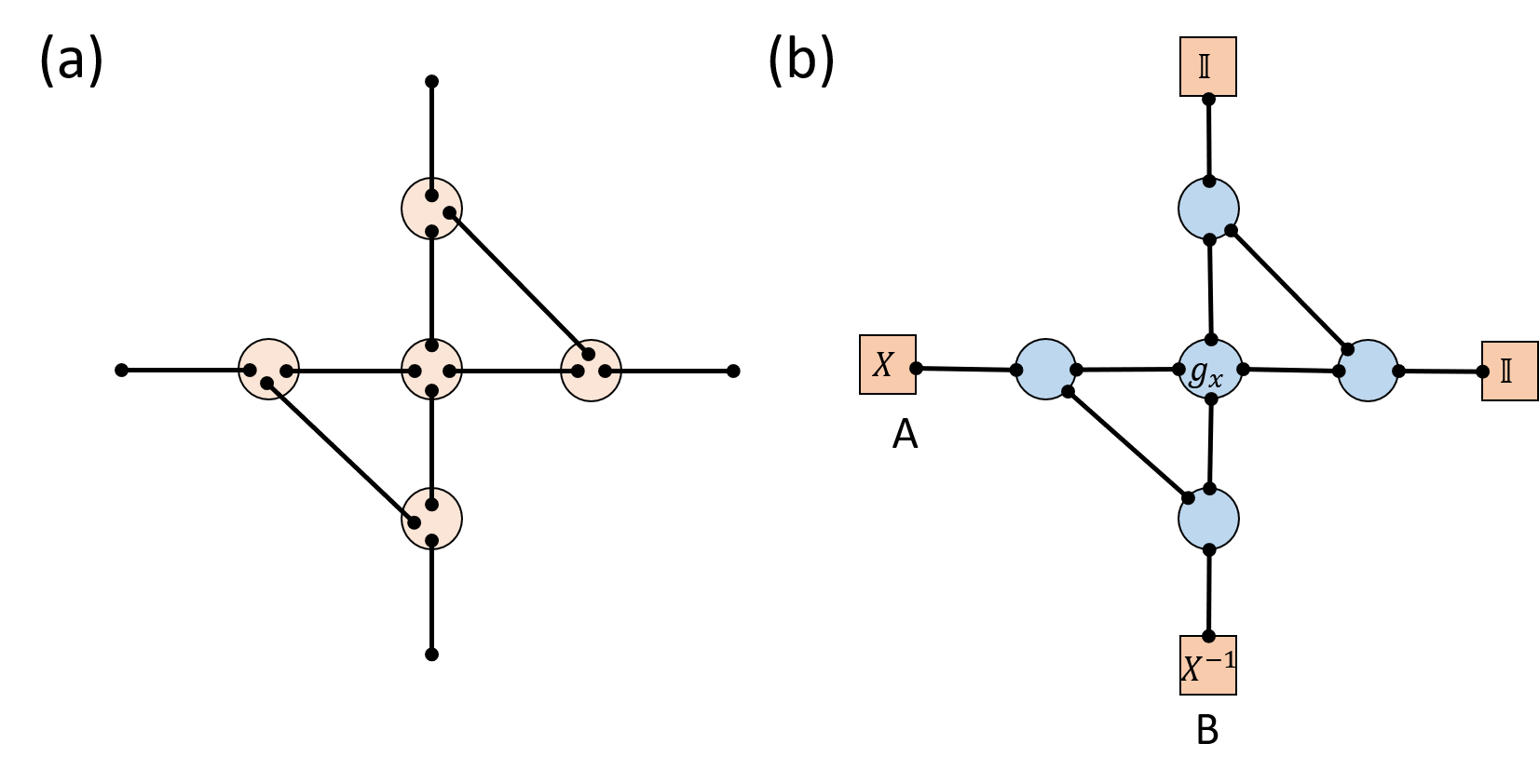}
\caption{(a)~Illustration of the random tensor network.
An EPR pair is defined for each link (line with dots), and a projection is applied to each vertex (circles).
(b)~The permutation spin model that computes the R\'enyi negativity of a random tensor network.
Each vertex carries a permutation group element $g_x\in S_k$, and the boundary condition is defined by $g_x=X$ on $A$, $g_x=X^{-1}$ on $B$ and $g_x=\mathbb{I}$ elsewhere, where $X$ is a $k$-cycle (see text). \label{fig:RTN}}
\end{figure}

To gain more intuition and insight on R\'enyi negativities $N_k$ in holographic duality, it is helpful to study those first in RTNs as a simpler toy model.
Our computation is based on the same technique as the R\'enyi entropy calculation in Ref.~\cite{hayden2016holographic}.
A similar calculation for~$k=3$ has been carried for random stabilizer tensor networks in Ref.~\cite{nezami2016multipartite}.
We also note that very sharp results are known for mixed states induced by a \emph{single} random tensor~\cite{aubrun2012partial,fukuda2013partial,banica2013asymptotic}; see also~\cite{shapourian2020entanglement} for recent developments.
In the following we give a heuristic derivation of our results for RTN.
We refer to \cref{app:permutations,app:domains,app:link,app:spec,app:spin model} for more details and rigorous proofs.

An RTN is a projected entangled pair state (PEPS)~\cite{verstraete2004valence}.
For a given graph with dangling edges, the corresponding RTN is defined by preparing an EPR pair or maximally entangled state~$\ket{L_{xy}}$ for each link $\overline{xy}$, and then projecting each vertex into a random state~$\ket{V_x}$:
\begin{align}\label{eq:RTN}
    \ket{\Psi}=\bigl(\otimes_x\bra{V_x}\bigr)\bigl(\otimes_{\overline{xy}}\ket{L_{xy}}\bigr),
\end{align}
as is shown in \cref{fig:RTN} (a).
If a vertex~$x$ has $p$ neighbors, there are $p$ qudits at $x$, and the state~$\ket{V_x}$ is chosen independently and uniformly from the Hilbert space of all $p$ qudits, which has total dimension~$D^p$.%
\footnote{More precisely, the random state $\ket{V_x}$ is defined as $\ket{V_x}=U\ket{0}$ with an arbitrary reference state $\ket{0}$ and a Haar random unitary $U$.}
The tensors~$\ket{V_x}$ on different sites are independent from each other.
The link states $\ket{L_{xy}}$ are entangled state of two qudits, in the Hilbert space of dimension $D^2$.
For simplicity we choose $\ket{L_{xy}}$ to be a maximally entangled state (but see the discussion at the end of this section).
After projecting on all bulk vertices, the RTN state~$\ket{\Psi}$ lives in the Hilbert space of the remaining boundary qudits at the end of the dangling edges of the graph.%
\footnote{Equivalently, one may add `boundary vertices' to each dangling edge to obtain a graph without dangling edges. This is the perspective taken in the appendix.}

Given the RTN state in \cref{eq:RTN}, we now choose two regions $A$ and $B$ on the boundary, and compute the associated R\'enyi negativity $N_k$ using \cref{eq:Nk2}:
\begin{align*}
    N_k=\frac{\bra{\Psi}^{\otimes k} P_A(X) P_B(X^{-1}) \ket{\Psi}^{\otimes k}}{\langle{\Psi}|{\Psi}\rangle^k}
    =\frac{\bra{L}^{\otimes k} P_A(X) P_B(X^{-1}) \prod_x \left(\ket{V_x}\bra{V_x}\right)^{\otimes k}\ket{L}^{\otimes k}}{\bra{L}^{\otimes k} \prod_x \left(\ket{V_x}\bra{V_x}\right)^{\otimes k}\ket{L}^{\otimes k}}
\end{align*}
with $\ket{L}=\otimes_{\overline{xy}}\ket{L_{xy}}$ and all identity operators omitted.
In principle, we have to compute the average of this random ratio.
However, the fluctuations of both the numerator and the denominator are suppressed in the large bond dimension limit.
Therefore,
\begin{align*}
    N_k \simeq \overline{N_k} \simeq \frac{Z_{k}^{AB}}{Z_k^{\emptyset}},
\end{align*}
where
\begin{align*}
    Z_k^{AB} &=  \bra{L}^{\otimes k} P_A(X) P_B(X^{-1}) \prod_x \overline{\left(\ket{V_x}\bra{V_x}\right)^{\otimes k}} \ket{L}^{\otimes k} \\
    Z_k^{\emptyset} &= \bra{L}^{\otimes k} \prod_x \overline{\left(\ket{V_x}\bra{V_x}\right)^{\otimes k}} \ket{L}^{\otimes k}.
\end{align*}
Here we used that the tensors $\ket{V_x}$ are independently chosen at random for each vertex.
This random average can be easily computed.
The key mathematical equality is the random average of projector on each site:
\begin{align}\label{eq:average1site}
    \overline{\ket{V_x}\bra{V_x}^{\otimes k}}=\frac1{C_k}\sum_{g\in S_k}P_x(g).
\end{align}
Here $g$ is a permutation element in permutation group $S_k$, and we recall that $P_x(g)$ denotes the corresponding permutation operator acting on the $k$ copies of site $x$.
$C_k$ is a normalization constant.
Using \cref{eq:average1site}, the random average of the numerator takes the form of the partition function in a classical statistical mechanics problem:
\begin{align*}
    Z_k^{AB} =\sum_{\left\{g_x\right\}}e^{-\mathcal{A}\left[\left\{g_x\right\}\right]}, \qquad
    e^{-\mathcal{A}\left[\left\{g_x\right\}\right]} \equiv \bra{L}^{\ot k} P_A(X) P_B(X^{-1}) \prod_x P_x(g_x) \ket{L}^{\ot k}
\end{align*}
and similar for $Z_k^{\emptyset}$.
This is a statistical mechanics model with discrete `spins'~$g_x$ summed over the $n!$ elements of the permutation group.
Due to the simple form of state $\ket{L}$ (a tensor product of EPR pairs for each link), the action $\mathcal{A}\left[\left\{g_{x}\right\}\right]$ has a simple form consisting of only two-body interaction terms. Explicitly we have
\begin{align}\label{eq:spin model action}
    \mathcal{A}\left[\left\{g_{x}\right\}\right]
=-\log(D) \sum_{\overline{xy}}\left(\chi\left(g_x^{-1}g_y\right)-k\right)
=\log(D) \sum_{\overline{xy}}d(g_x, g_y)
\end{align}
up to an overall constant.
Here, $\chi(g)$ is the number of cycles in a permutation~$g$ (which is~$k$ for the identity and~$1$ for a $k$-cycle).
If we write $d(g,h)$ for the minimal number of swaps to go from $g$ to $h$ (the so-called \emph{Cayley metric} on the permutation group $S_k$) then we have $d(g,h) = k-\chi(g^{-1}h)$.
The cyclic permutations $P_A(X)$ and $P_B(X^{-1})$ (and the identity acting on the complement of $AB$) play the role of the boundary condition for the classical spin model.
See \cref{fig:RTN}~(b) for an illustration.

\begin{figure}[t]
\centering
\includegraphics[width=3.3in]{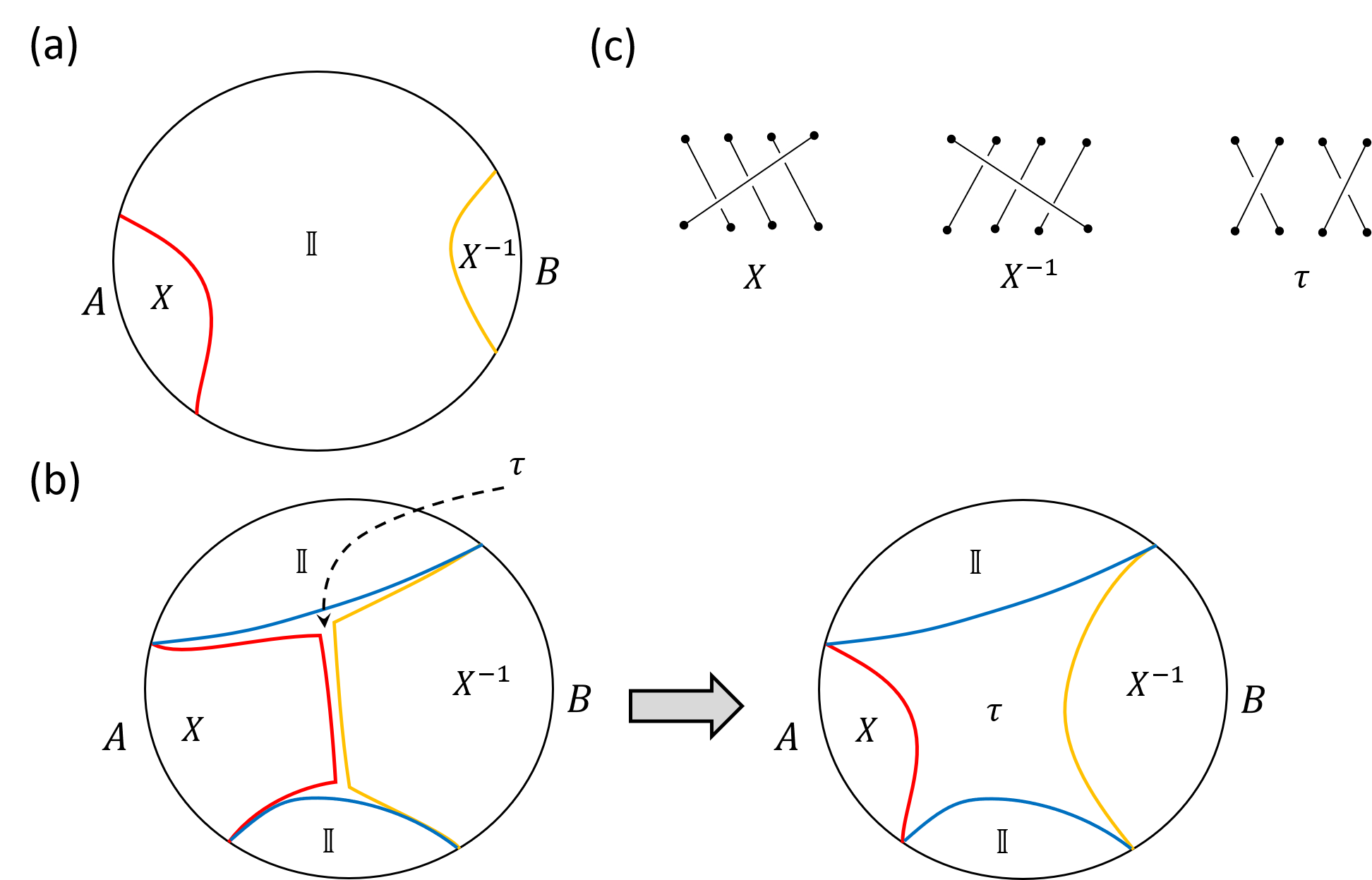}
\caption{The domain configuration in the case that (a) $I_{RT}(A:B)=0$ and (b) $I_{RT}(A:B)>0$. The red and orange curves are RT surfaces of $A$ and $B$, respectively, while the blue curves in (b) are the RT surface for $AB$. The RT surfaces are domain boundaries, with the value of $g_x$ in each domain indicated on the figure. (c) Illustration of permutation $X, X^{-1}$ and an example of $\tau$ in (b). There are multiple $\tau$ which contribute identically to the action (see text).\label{fig:domain}}
\end{figure}

Since the action $\mathcal{A}\left[\left\{g_{x}\right\}\right]$ prefers neighboring spins to be parallel (since $d$ is a metric; equivalently, $g_x=g_y$ maximizes $\chi(g_x^{-1}g_y)$), at large $D$ (strong coupling, or low temperature) the dominant configurations contain large domains of spins with a small free energy cost given by domain walls.
For the boundary condition we are concerned about here, the lowest action configuration turns out to take the form illustrated in \cref{fig:domain}.
There are two basic cases to consider:

For regions $A$, $B$ such that RT surface of $AB$ is the disjoint union of the individual RT surfaces, $\gamma_{AB} = \gamma_A \cup \gamma_B$, the domain configuration that minimizes the action contains three domains, which are filled with $X$, $X^{-1}$, and $\mathbb{I}$, respectively, and separated by the minimal surfaces~$\gamma_A$ and~$\gamma_B$.%
\footnote{Here and in the following we assume that the RT surface for a given boundary region is unique.}
Note that in this case the mutual information is $I_{RT}(A:B)=0$ when calculated using the Ryu-Takayanagi formula.

When the RT surface for $AB$ is different from the union of the individual RT surfaces, i.e., when the mutual information $I_{RT}(A:B)>0$, the three-domain configuration is suppressed, and the dominant configuration becomes the four-domain configuration shown in \cref{fig:domain}~(b).
Here, the $X$ and $X^{-1}$ domains are still bounded by $\gamma_A$ and $\gamma_B$, respectively, but now the identity domain is bounded by $\gamma_{AB}$.
The fourth domain (we assume for simplicity that it is connected) fills the rest of the bulk with a permutation element $\tau$.
To understand what are dominant configurations, we can first think of a configuration with three domains $\mathbb{I},X,X^{-1}$ as is illustrated in \cref{fig:domain}~(b).
Inserting a small fourth domain $\tau$ corresponds to splitting the domain walls into two.
For generic $\tau$, such as splitting leads to an `energy cost' in the statistical model.
For example, if we split a domain wall between $X$ and $\mathbb{I}$ into two domain walls from $X$ to $\tau$ and then to $\mathbb{I}$, the tension per bond of the domain wall becomes $d(X,\tau)+d(\tau,\mathbb{I})$ which is generally bigger or equal to $d(X,\mathbb{I})$.
Here $d(g,h)$ is the distance in the permutation group $S_k$ introduced in \cref{eq:spin model action}.
Such a splitting has zero energy cost only if $d(X,\tau)+d(\tau,\mathbb{I})=d(X,\mathbb{I})$.
Therefore to require all domain walls can split in the same way we need the following requirements:
\begin{equation}\label{eq:taucondition}
\begin{aligned}
    d(X,\tau)+d(\tau,\mathbb{I})&=d(X,\mathbb{I}), \\*
    d(X^{-1},\tau)+d(\tau,\mathbb{I})&=d(X^{-1},\mathbb{I}), \\*
    d(X,\tau)+d(\tau,X^{-1})&=d(X,X^{-1}).
\end{aligned}
\end{equation}
If these conditions are satisfied, the three domain configuration will split (since there is no cost), and after splitting each domain wall will deform into minimal surfaces, as is illustrated in \cref{fig:domain}.
For each $k$, there is a finite number of permutations $\tau\in S_k$ which satisfy the conditions~\eqref{eq:taucondition}, an example of which is given in \cref{fig:domain}~(c).
In general, the $\tau$ are non-crossing pairings.
We define and discuss this in detail in \cref{app:permutations,app:domains}.

In the limit of large bond dimension~$D$, the dominant configuration determines the behavior of $N_k$.
In the first case (\cref{fig:domain} (a)), we have
\be\la{eq:logNdis}
\log N_k \simeq -(k-1)\log(D)\left(\left|\gamma_A\right|+\left|\gamma_B\right|\right) \simeq -(k-1) \, S^{(k)}(\rho_{AB}),
\ee
hence is determined by the $k$-th R\'enyi entropy of region~$AB$.
This result implies that the partial transpose has trivial effect, which is intuitive, since in \cref{fig:domain} (a) the action stays the same if we change the $X^{-1}$ domain to $X$, which is the configuration that determines $S^{(k)}_{AB}$~\cite{hayden2016holographic}.

We now focus on the nontrivial case shown in \cref{fig:domain}~(b), when regions $A,B$ have nontrivial Ryu-Takayanagi mutual information.
In this case, the tension of domain wall (i.e., action cost for each link crossing the domain wall) depends on the parity of~$k$.
We will write down the results here and leave more details of the derivation in \cref{app:domains}.
For a connected $\tau$-domain:
\begin{align}
    \log N_{2n-1}^\odd &\simeq-(n-1)\log(D)\left(\left|\gamma_{AB}\right|+\left|\gamma_A\right|+\left|\gamma_B\right|\right)
    ~-~\log a_{2n-1},\label{eq:logNodd}\\
    \log N_{2n}^\even &\simeq-n\log(D)\left|\gamma_{AB}\right|-(n-1)\log(D)\left(\left|\gamma_A\right|+\left|\gamma_B\right|\right)
    ~-~\log a_{2n}.\label{eq:logNeven}
\end{align}
The leading order terms ($\propto \log D$) are proportional to the domain wall area, with the coefficient determined by the tension of the domain wall, which is $d(X,\tau)$, $d(X^{-1},\tau)$, and $d(I,\tau)$ for $\gamma_A,\gamma_B$, and $\gamma_{AB}$, respectively.
The last term $\log a_k$ is a finite correction in the large $D$ limit, with $a_k$ the number of $\tau$ that gives this leading contribution to $\log N_k$. The explicit form of $a_k$ is given in \cref{app:permutations}.

With the R\'enyi negativities computed, we can consider their analytic continuation.
Here we proceed heuristically and simply take the analytic continuation of the formulas derived above, while deferring rigorous proofs to the appendix.
As we discussed earlier, the logarithmic negativity is obtained by analytically continuing $\log N_{2n}$ to $n\rightarrow \frac12$, which leads to
\begin{align}\label{eq:negativityasymptotics}
    E_N(\rho_{AB}) \simeq \frac{\log D}2\left(\left|\gamma_A\right|+\left|\gamma_B\right|-\left|\gamma_{AB}\right|\right) + \log \frac{8}{3\pi} + o(1) \simeq \frac12I(A:B) + \log \frac{8}{3\pi}.
\end{align}
The leading order term $\propto \log D$ is half of the mutual information between the two regions.
The constant correction term $\log\frac{8}{3\pi}<0$ reflects the fact that the entanglement in~$\rho_{AB}$ is not exactly pure bipartite.
We give a precise derivation of \cref{eq:negativityasymptotics} in \cref{app:spec}, where we compute more generally the \emph{eigenvalue distribution} of the partial transpose in the large bond dimension limit.

To state our result on the eigenvalue distribution of the partial transpose, consider the empirical eigenvalue distribution
\begin{align}\label{eq:empirical eigs}
  \mu_D = \frac1{D^{\lvert\gamma_A\rvert + \lvert\gamma_B\rvert}} \sum_{i=1}^{D^{\lvert\gamma_A\rvert + \lvert\gamma_B\rvert}} \delta_{s_i},
\end{align}
where $s_1 \geq s_2 \geq \dots$ denote the eigenvalues of $M_{AB} = D^{\frac12(\lvert\gamma_A\rvert+\lvert\gamma_B\rvert+\lvert\gamma_{AB}\rvert)} \rho_{AB}^{T_B}$ (a suitable rescaling of the partial transpose) and $\delta_s$ denotes the Dirac measure.
We note that $\rk \rho_{AB}^{T_B} = \rk M_{AB} \leq D^{\lvert\gamma_A\rvert + \lvert\gamma_B\rvert}$, so $\mu_D$ captures all nonzero eigenvalues of the partial transpose.
Then the $\mu_D$ are a sequence of random probability measures, and we prove that for $D\to\infty$ they converge weakly, in probability, to the \emph{Wigner semicircle distribution}
\begin{align}
    d\mu_W = \frac1{2\pi} \sqrt{4-\lambda^2} \, d\lambda
\end{align}
on~$[-2,2]$.
This result implies \cref{eq:negativityasymptotics} at once.

We emphasize that the formulas given above apply when the $\tau$-domain (that is, the domain bounded by $\gamma_A$, $\gamma_B$, and $\gamma_{AB}$) is connected, as in \cref{fig:domain}~(b).
In general, we can choose a different permutation~$\tau$ for each connected component.
We give general formulas in \cref{app:domains,app:spec}.
In particular, we find that the additive correction in \cref{eq:negativityasymptotics} is in general $\log\frac{8}{3\pi}$ times the number of connected components in the $\tau$ domain.

\subsection{Non-maximally entangled link states}\label{sec:nmels}
Here we would like to discuss a more general situation when the link state~$\ket{L_{xy}}$ in RTN is not maximally entangled.
This discussion is motivated by the comparison between RTN and AdS/CFT.
In a simple RTN with maximally entangled link states, in the large bond dimension limit all integer R\'enyi entropies have the same value and are therefore equal to the von Neumann entropy (assuming the RT surface is unique).
In contrast, the R\'enyi entropies in a CFT have a nontrivial dependence on the R\'enyi parameter~\cite{Lewkowycz:2013nqa,dong2016gravity}, which shows that the entanglement spectrum (i.e., the eigenvalue spectrum of the reduced density matrix~$\rho_A$ of a region $A$) is nontrivial.
The RTN with non-maximally entangled link states provides a toy model of systems with an $n$-dependent R\'enyi entropy, although there are still differences with the AdS/CFT case (see~\cite{hayden2016holographic,Dong:2018seb,Akers:2018fow} for details).

For simplicity we assume each link carries the same state $\ket{L_{xy}}$, which however now is no longer maximally entangled.
Denote the reduced density matrix of either subsystem of the single link state as~$\rho_e$.
This corresponds to modifying the spin model action in \cref{eq:spin model action} to
\begin{align}\label{eq:action nonmaximal}
    \mathcal{A}\left[\left\{g_x\right\}\right]=\sum_{\overline{xy}}J\left(g_x^{-1}g_y\right), \qquad
    J(h)=-\log {\rm tr}\left(P(h)\rho_e^{\otimes k}\right).
\end{align}
More explicitly, if $h$ contains cyclic permutations of length $k_1,k_2,...,k_c$, we have
\begin{align}
    J(h)=\sum_{i=1}^c\left(k_i-1\right)S_{k_i}\left(\rho_e\right).
\end{align}
For a maximally entangled link state, $\rho=\frac1D\mathbb{I}$ and $J(h)$ reduces to \cref{eq:spin model action}.
In the case of a non-maximally entangled link state, however, the permutations~$\tau$ that satisfy \cref{eq:taucondition} no longer contribute equally to $N_k$.
In \cref{app:link} we present more detailed discussion which shows that, while all these $\tau$ contribute the same way to the $\mathbb{I}$\,--\,$\tau$ domain wall, the contribution to the $X$--\,$\tau$ and $X^{-1}$\,--\,$\tau$ domain walls are different.
The dominant contribution is given by the $\tau$ such that $X\tau$ and $X^{-1}\tau$ has smallest number of nontrivial cycles.
These are the `neighboring pair' permutation such as $\tau=(12)(34)$ illustrated in \cref{fig:domain}~(c).
There are two such permutations for even $k$, and $k$ such permutations for odd $k$.%
\footnote{However, our calculation does not exclude the possibility that, in the case of non-maximally entangled link states, other permutations could in principle be even more dominant than these non-crossing pairing $\tau$'s. Indeed, this can be the case near the mutual information transition.}
This motivates us to focus on these permutations in the gravity discussion in next section.

\section{Negativity in holographic duality}\label{sec:holo}
In this section, we study the R\'enyi negativity $N_k$ and its various limits for general regions $A$, $B$ in holographic theories.
Starting with the case of integer $k$, we may rewrite \cref{eq:Nk2} as
\be\la{nkr}
N_k = \fr{Z_k}{Z_1^k},
\ee
where $Z_k$ is the boundary partition function on a $k$-fold branched cover~$M_k^{A,B}$ obtained by gluing $k$ copies of the original boundary spacetime $M_1$ cyclically along $A$ and anti-cyclically along $B$.%
\footnote{For the R\'enyi entropy of $\r_{AB}$, we would instead glue cyclically along both $A$ and $B$.}
This $k$-fold cover of the boundary spacetime enjoys a manifest $\bZ_k$ replica symmetry generated by the cyclic permutation of the $k$ copies.

We will use the holographic duality to calculate~$Z_k$ to leading order in the gravitational constant~$G$:
\be\la{zkb}
Z_k = e^{-I[\cB_k]}
\ee
Here $I[\cB_k]$ is the on-shell action of an appropriate bulk saddle point solution $\cB_k$ whose asymptotic boundary is the $k$-fold cover $M_k^{A,B}$.
In case there are multiple saddle points satisfying this boundary condition, we should choose the dominant saddle, that is, the one with the smallest on-shell action.

In general, the bulk saddle points will not obey the $\bZ_k$ replica symmetry on the boundary.
Some of them do, but many do not.
In the following subsections, we will study first the contributions of replica symmetric saddle points, and then those of replica nonsymmetric saddles.
As we will see, an interesting feature of the R\'enyi negativity is that replica symmetric saddles do not always dominate the replicated partition function $Z_k$.

\subsection{Replica symmetric saddle}\la{ssecsym}
We now calculate the contribution of a replica symmetric saddle $\cB_k^\sym$ to the R\'enyi negativity.
This discussion applies to both even and odd $k$.

\begin{figure}[t]
\centering
\includegraphics[width=\textwidth]{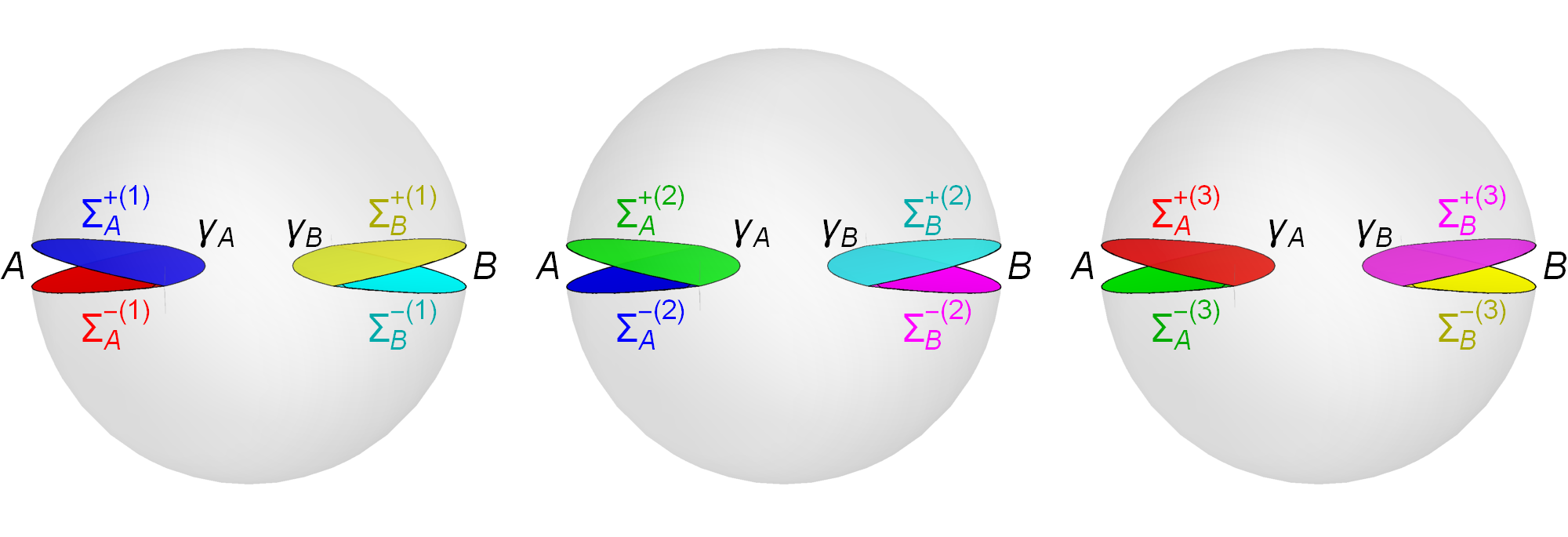}
\caption{The topology of a replica symmetric saddle contributing to the holographic R\'enyi negativity~$N_k$ for~$k=3$.
The three replicas are first cut along $\S_A$, $\S_B$, and then the cut surfaces $\S_A^{\pm(i)}$, $\S_B^{\pm(i)}$, $i=1,\dots,3$ with the same color are glued together (cyclically along $\S_A$ and anti-cyclically along $\S_B$).}
\label{figrs}
\end{figure}

To preserve the $\bZ_k$ symmetry in the bulk, we simply extend the replica construction of the $k$-fold cover $M_k^{A,B}$ topologically into the bulk.
In other words, we consider the bulk topology obtained by starting with $k$ copies of the original bulk spacetime~$\cB_1$, cutting each of them along two bulk codimension-1 surfaces $\S_A$, $\S_B$, and gluing these $k$ pieces together cyclically along $\S_A$ and anti-cyclically along $\S_B$ (\cref{figrs}).
The surface~$\S_A$ is required to have boundary $A \cup \g_A$, where $\g_A$ is a bulk codimension-2 surface homologous to $A$; similarly, the boundary of $\S_B$ is $B\cup \g_B$.
Once we have fixed this bulk topology, we impose the equations of motion and find a replica symmetric saddle $\cB_k^\sym$.
The codimension-2 surfaces $\g_A$, $\g_B$ become analogues of the RT surfaces for $A$, $B$, and the codimension-1 surfaces $\S_A$, $\S_B$ may be called homology hypersurfaces.

Due to its replica symmetry, we may take a quotient of~$\cB_k^\sym$ by the group~$\bZ_k$ and write its contribution to the replicated partition function \er{zkb} as
\be\la{zksym}
Z_k^\sym = e^{-I[\cB_k^\sym]} = e^{-k I[\hat\cB_k^\sym]},
\ee
where $\hat\cB_k^\sym \eq \cB_k^\sym \!/ \bZ_k$ is the quotient space, following~\cite{Lewkowycz:2013nqa,dong2016gravity}.
The quotient space is generally an orbifold with codimension-2 conical singularities along the fixed points of the $\bZ_k$ replica symmetry.
These fixed points may be identified with $\g_A \cup \g_B$, and the opening angle of the conical defects is $\frac{2\pi}k$.
By construction, the on-shell action $I[\hat\cB_k^\sym]$ of the quotient space does not include any localized contribution from (or boundary term on) the conical singularity.

The quotient space $\hat\cB_k^\sym$ may alternatively be obtained by finding a bulk solution to the equations of motion with the asymptotic boundary being the original boundary spacetime $M_1$, subject to the additional boundary condition that there are conical defects on two codimension-2 surfaces $\g_A$ and $\g_B$ (homologous to $A$ and $B$, respectively) with opening angle~$\frac{2\pi}k$.%
\footnote{We may enforce the boundary condition by including two cosmic branes on $\g_A$, $\g_B$, with tension~$\frac{k-1}{4k G}$~\cite{Lewkowycz:2013nqa,dong2016gravity}.}
The advantage of this alternative construction is that it applies to non-integer $k$ and provides the analytic continuation of the quotient space away from integer~$k$.
Instead of calling this solution $\hat\cB_k^\sym$, we will denote it by $\cB\(M_1, \g_A^{(k)}, \g_B^{(k)}\)$ to emphasize its boundary conditions, where $\g_A^{(k)}$, $\g_B^{(k)}$ indicate conical defects with opening angle $\frac{2\pi}k$.
We write its on-shell action simply as $I\(M_1, \g_A^{(k)}, \g_B^{(k)}\)$, and therefore \cref{zksym} becomes
\begin{equation*}
Z_k^\sym = e^{-k I\(M_1, \g_A^{(k)}, \g_B^{(k)}\)}.
\end{equation*}
If this replica symmetric saddle makes the dominant contribution to the replicated partition function $Z_k$, the R\'enyi negativity \er{nkr} would therefore be given by
\be\la{nksym}
\log N_k^\sym = -k \[I\(M_1, \g_A^{(k)}, \g_B^{(k)}\) - I\(M_1\)\],
\ee
where $I\(M_1\)$ denotes the on-shell action of the smooth bulk saddle $\cB_1$ with asymptotic boundary $M_1$ (and no conical defect).
The superscript indicates that this would be the value of the R\'enyi negativity \emph{assuming the dominant saddle is the replica symmetric one}, which, as mentioned, need not be the case.

Assuming replica symmetry, the analytic continuation in $k$ is therefore be the same for even and odd $k$, and given explicitly by \cref{nksym}.
In particular, the logarithmic negativity \er{en} vanishes at order~$O(G^{-1})$:
\begin{equation*}
E_N^\sym
= \lim_{k\to1} \log N_k^\sym = 0.
\end{equation*}
Furthermore, the refined R\'enyi negativity are in both the odd case~\er{sto} and the even case~\er{ste} given by
\be\la{stsym}
S_{AB}^{T_B,(k,\text{sym})} = -k^2 \pa_k \(\fr{1}{k} \log N_{k}^\sym\) = k^2 \pa_k I\(M_1, \g_A^{(k)}, \g_B^{(k)}\) = \fr{|\g_A^{(k)} \cup \g_B^{(k)}: M_1|}{4G},
\ee
where $|\g_A^{(k)} \cup \g_B^{(k)}: M_1|$ denotes the area of the conical defect $\g_A^{(k)} \cup \g_B^{(k)}$ with opening angle $\frac{2\pi}k$ in the bulk solution with asymptotic boundary $M_1$.
The last equality follows from the same variational argument as in~\cite{Lewkowycz:2013nqa,dong2016gravity}.
Taking the $k\to1$ limit of \cref{stsym}, we then find following expression for the partially transposed entropy defined in \cref{st}:
\be\la{st1sym}
S_{AB}^{T_B,\sym} = \fr{|\g_A^{(1)} \cup \g_B^{(1)}: M_1|}{4G} = \fr{|\g_A^\ext|+|\g_B^\ext|}{4G},
\ee
where we have used the fact shown in~\cite{Lewkowycz:2013nqa} that conical defects becomes extremal surfaces in the limit where the opening angle approaches $2\pi$, and $|\g_A^\ext|$, $|\g_B^\ext|$ denote the areas of the extremal surfaces $\g_A^\ext$, $\g_B^\ext$ in the original bulk solution $\cB_1$.
Taking instead the $k\to2$ limit of \cref{stsym}, we see that the refined R\'enyi-2 negativity defined in \cref{st2} are given by:
\be\la{st2sym}
S_{AB}^{T_B(2,\text{sym})} = \fr{|\g_A^{(2)} \cup \g_B^{(2)}: M_1|}{4G}.
\ee

In general, the conical bulk solution $\cB\(M_1, \g_A^{(k)}, \g_B^{(k)}\)$ is difficult to obtain explicitly due to nontrivial gravitational backreaction from the conical defects.
However, it can easily be obtained in \emph{fixed-area states}~\cite{Dong:2018seb,Akers:2018fow,Dong:2019piw} where the areas of $\g_A$ and $\g_B$ are fixed to some particular values, which we will denote by $|\g_A|$ and $|\g_B|$.
In such states, the gravitational backreaction from the conical defects is frozen, and the on-shell action of the bulk solution $\cB\(M_1, \g_A^{(k)}, \g_B^{(k)}\)$ is simply
\begin{equation*}
I\(M_1, \g_A^{(k)}, \g_B^{(k)}\) = I(M_1) + \(1- \fr{1}{k}\) \fr{|\g_A|+|\g_B|}{4G},
\end{equation*}
following a similar calculation as in~\cite{Dong:2018seb,Akers:2018fow,Dong:2019piw}.
Therefore, the R\'enyi negativity~\er{nksym} simplifies to
\be\la{nsfa}
\log N_k^\sym = -\frac{k-1}{4G} \left( |\g_A|+|\g_B| \right),
\ee
which matches the RTN result~\er{eq:logNdis} in the zero mutual information case if we identify $\log D = \frac1{4G}$.
Moreover, the refined R\'enyi negativities~\er{stsym} become independent of $k$:
\begin{align*}
S_{AB}^{T_B(k,\text{sym})} = \fr{|\g_A|+|\g_B|}{4G} = S_A + S_B,
\end{align*}
which is then also the value of the partially transposed entropy~\er{st1sym} and the refined R\'enyi-2 negativity~\er{st2sym}.

As mentioned earlier, the replica symmetric saddle discussed here is not always the dominant one.
We will show this explicitly in the next two subsections by constructing replica nonsymmetric saddles.

\subsection{Replica nonsymmetric saddle for even \texorpdfstring{$k$}{k}}
We now study the contributions of bulk saddles that break the $\bZ_k$ replica symmetry of the boundary $k$-fold cover $M_k^{A,B}$.
In this subsection, we focus on the case of even $k=2n$.

To obtain a replica nonsymmetric saddle with asymptotic boundary $M_{2n}^{A,B}$, we use a different cutting and gluing procedure from the one used in the previous subsection.
Starting again with $2n$ copies of the original bulk spacetime $\cB_1$, we cut each of them along \textit{three} non-overlapping codimension-1 surfaces $\S_A$, $\S_B$, and $\S_{\ol{AB}}$.
Analogous to $\S_A$ and $\S_B$ that appeared previously, the surface $\S_{\ol{AB}}$ is required to have boundary $\ol{AB} \cup \g_{AB}$ where $\g_{AB}$ is a bulk codimension-2 surface that is then also homologous to~$AB$.%
\footnote{Without loss of generality, we assume that the entire system is in a pure state, and therefore the codimension-2 surface $\g_{AB}$ may also be called $\g_{\ol{AB}}$.}
To make the construction nontrivial, we assume that $\g_{AB}$ is different from $\g_A \cup \g_B$.

\begin{figure}[t]
\centering
\includegraphics[width=\textwidth]{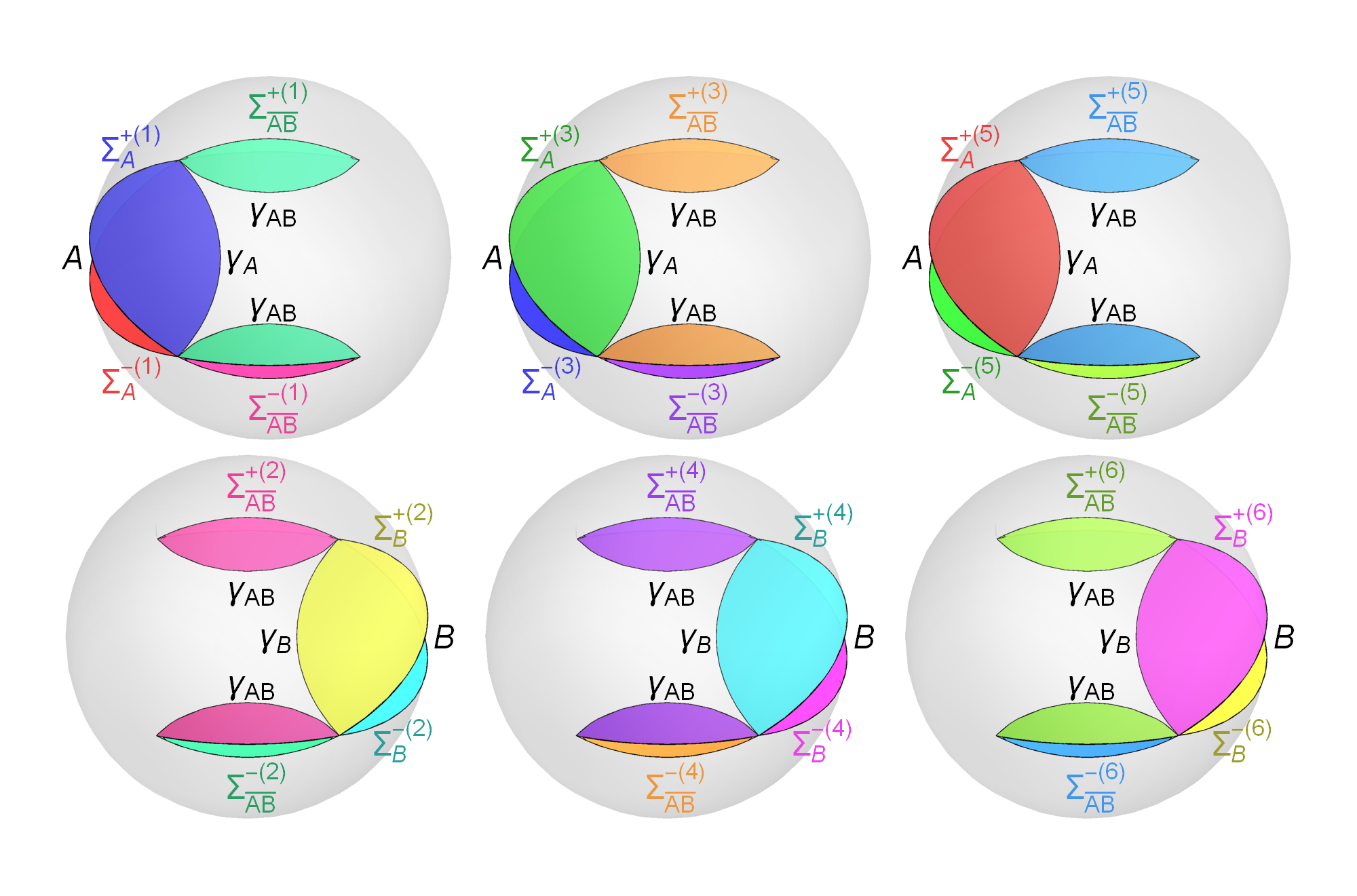}
\caption{The topology of a replica nonsymmetric saddle contributing to the holographic R\'enyi negativity~$N_k$ for~$k=6$.
The 6 copies are first cut along $\S_A$, $\S_B$, $\S_{\ol{AB}}$, and then the cut surfaces $\S_A^{\pm(i)}$, $\S_B^{\pm(i)}$, and $\S_{\ol{AB}}^{\pm(i)}$ with the same color are glued together (cyclically along $\S_A$ in copies $1$, $3$, $5$, anti-cyclically along $\S_B$ in copies $2$, $4$, $6$, and pairwise along $\S_{\ol{AB}}$ between copies $2j-1$ and $2j$ for $j=1,\dots,3$).
Note that the surface $\S_{\ol{AB}}$ shown here has two connected components on each copy.}
\label{figeven}
\end{figure}

The gluing procedure is determined by specifying a permutation of the $2n$ copies for each of the three cuts $\S_A$, $\S_B$, and $\S_{\ol{AB}}$.
We will focus on the following particularly simple prescription (illustrated in \cref{figeven}).
Let us label the copies by $1,2,\dots,2n$.
Along $\S_A$, we glue the $n$ copies labeled by the odd integers $1,3,\ldots,2n-1$ cyclically, and we glue each of the remaining copies to itself.
Along $\S_B$, we glue the $n$ copies labeled by the even integers $2,4,\ldots,2n$ anti-cyclically, and we glue each of the remaining copies to itself.
Along $\S_{\ol{AB}}$, we glue the first two copies together, then the next two copies, and so on, until we have glued the last two copies.
This gluing procedure manifestly breaks the full $\bZ_{2n}$ replica symmetry, although it preserves a $\bZ_n$ subgroup that is generated by $(1~3~\dots~2n\!-\!1)(2~4~\dots~2n)$, which cyclically permutes the $n$ odd copies and the $n$ even copies separately.

It is not difficult to verify that the cutting and gluing procedure described above indeed leads to a bulk manifold with asymptotic boundary $M_{2n}^{A,B}$.
To see this, we note that the boundary entangling surface $\pa A$ is asymptotically approached by both $\g_{A}$ and $\g_{AB}$, and therefore the boundary permutation around $\pa A$ (which determines the gluing procedure along $A$ on the boundary) is the composition of two permutations:
the one around $\g_{AB}$ which is $\tau = (12)(34)\cd(2n\!-\!1 ~ 2n)$, followed by the one around $\g_A$ which is $X \tau = (1~3~\dots~2n\!-\!1)$.
This composition is the full $2n$-cycle $X = (1~2~\dots~2n)$, which precisely what is needed to glue all $2n$ boundary copies cyclically along $A$.
Similarly, the boundary permutation around~$\pa B$ is the composition of the permutation $\tau$ around $\g_{AB}$ followed by the permutation $X^{-1} \tau = (2n ~ 2n\!-\!2 ~ \dots ~ 2)$ around $\g_B$.
This composition is $X^{-1} = (2n ~ 2n\!-\!1 \dots 1)$, correctly gluing all $2n$ boundary copies anti-cyclically along $B$.
Thus we see that even though the gluing procedure in the bulk manifestly breaks the $\bZ_{2n}$ replica symmetry, it is preserved on the boundary.

The bulk topology resulting from the cutting and gluing procedure described above can also be obtained by an alternative, but equivalent, cutting and gluing procedure that is motivated by our RTN discussion in \cref{sec:rtn}.  In this alternative procedure, instead of cutting the copies along $\S_{\ol{AB}}$ we cut them along the ``middle surface'' $\S_{\text{mid}}$ (in addition to $\S_A$ and $\S_B$), where $\S_{\text{mid}}$ is the complement of $\S_A \cup \S_B \cup \S_{\ol{AB}}$ on the codimension-1 time slice.  The middle surface $\S_{\text{mid}}$ has the boundary $\g_A\cup \g_B \cup \g_{AB}$ and is the analogue of the $\tau$-domain in our RTN discussion.  In the alternative gluing procedure, we glue the copies along $\S_{\text{mid}}$ according to the permutation $\tau = (12)(34)\cd(2n\!-\!1 ~ 2n)$, along $\S_A$ according to $X = (1~2~\dots~2n)$, and along $\S_B$ according to $X^{-1} = (2n ~ 2n\!-\!1 \dots 1)$.  This alternative cutting and gluing procedure is equivalent to the one described in the previous paragraphs, as they become identical under a relabeling of the copies that has the effect of right-multiplying by $\tau$ the four permutations used in gluing along $\S_A$, $\S_B$, $\S_{\ol{AB}}$, and $\S_{\text{mid}}$.  The comments in this paragraph also apply to \cref{sec:hodd} where we discuss the case of odd $k$. In the following discussion, we will use the original cutting and gluing procedure described in the previous paragraphs, for it makes the nature of the conical defects described below more manifest.

As before, we impose the equations of motions after fixing the bulk topology using the cutting and gluing procedure.
We will call this replica nonsymmetric saddle $\cB_{2n}^\nsym$.
As mentioned above, this saddle breaks the full $\bZ_{2n}$ replica group down to a $\bZ_n$ subgroup generated by the cyclic permutation of the $n$ pairs of copies.
Therefore, it is useful to define the quotient space $\hat\cB_{2n}^\nsym \eq \cB_{2n}^\nsym \! / \bZ_n$, whose asymptotic boundary is $M_2^{AB}$, a $2$-fold cover%
\footnote{Note that $M_k^{AB}$ is in general different from $M_k^{A,B}$ because the $k$ copies are glued along $B$ differently.
The former is used for calculating the R\'enyi entropy, whereas the latter is for the R\'enyi negativity.
However, the two covering spaces are identical for $k=2$.}
of the original boundary $M_1$ branched over $AB$.
The quotient space has conical defects at the fixed point loci $\g_{A_1}^{(n)}$ and $\g_{B_2}^{(n)}$ with opening angle $\frac{2\pi}n$.
Here the subscripts $1$ and $2$ mean that the surface $\g_{A_1}^{(n)}$ is homologous to the first copy of the boundary region~$A$ in the $2$-fold cover~$M_2^{AB}$, while $\g_{B_2}^{(n)}$ is homologous to the second copy of~$B$.

The quotient space $\hat\cB_{2n}^\nsym$ may again be obtained alternatively as the bulk solution with asymptotic boundary $M_2^{AB}$ and with conical defects on two codimension-2 surfaces $\g_{A_1}^{(n)}$ and $\g_{B_2}^{(n)}$ (homologous to the first copy of region $A$ and to the second copy of region $B$, respectively) with opening angle $\frac{2\pi}n$.
This alternative construction, which we will call $\cB\(M_2^{AB}, \g_{A_1}^{(n)}, \g_{B_2}^{(n)}\)$, provides an analytic continuation of the quotient space away from integer~$n$.
We write its on-shell action as $I\(M_2^{AB}, \g_{A_1}^{(n)}, \g_{B_2}^{(n)}\)$.
Its contribution to the partition function~\er{zkb} is
\begin{align*}
Z_{2n}^{(\text{even,nsym})} = e^{-n I\(M_2^{AB}, \g_{A_1}^{(n)}, \g_{B_2}^{(n)}\)},
\end{align*}
where we have included ``even'' in the superscript to emphasize that this formula gives the even analytic continuation in $n$.

If the saddle thus constructed is the dominant one%
\footnote{There are additional replica nonsymmetric saddles with topologies obtained from cutting and gluing procedures dictated by different $\tau$ permutations.  One of them corresponds to $\tau = (23)(45)\cd(2n ~ 1)$ and is equivalent to the $\tau = (12)(34)\cd(2n\!-\!1 ~ 2n)$ saddle studied here.  The others do not preserve the residual $\bZ_n$ replica symmetry and are more difficult to analyze.  Whether or not they could dominate over the saddle studied here is an interesting question that we leave to future work.  In fixed-area states to be discussed momentarily, the additional saddles obtained from non-crossing pairings $\t$ have equal contribution to the saddle studied here at leading order in $G$.  The saddle studied here is motivated by what we found in RTNs in \cref{sec:nmels}: while in RTNs with maximally entangled link states all non-crossing pairings give equal and dominant contributions, in RTNs with non-maximally entangled link states those equivalent to the $\t$ studied here dominate over all other non-crossing pairings.  These comments also apply to our discussion of odd $k$ in \cref{sec:hodd}.}%
, the even analytic continuation of the R\'enyi negativity~\er{nkr} would be
\be\la{nke}
\log N_{2n}^{(\text{even,nsym})} = -n \[I\(M_2^{AB}, \g_{A_1}^{(n)}, \g_{B_2}^{(n)}\) -2 I\(M_1\)\],
\ee
where $I\(M_1\)$ again denotes the on-shell action of the original bulk saddle $\cB_1$.
Taking the $n\to \frac12$ limit, we find that the logarithmic negativity~\er{en} is
\be\la{ene}
E_N^\nsym = \lim_{n\to 1/2} \log N_{2n}^{(\text{even,nsym})} = I(M_1) - \fr{1}{2} I\(M_2^{AB}, \g_{A_1}^{(\fr{1}{2})}, \g_{B_2}^{(\fr{1}{2})}\).
\ee
This is a concrete expression of the logarithmic negativity in terms of 
the on-shell action of the bulk solution with asymptotic boundary $M_2^{AB}$ and with conical defects on $\g_{A_1}$, $\g_{B_2}$ (homologous to the first copy of region $A$ and to the second copy of region $B$, respectively) with opening angle $4\pi$.
In the special case where $A$ and $B$ are adjacent intervals in the vacuum state of a 2-dimensional CFT, the result is completely fixed by the conformal symmetry~\cite{Calabrese:2012ew} and agrees with \cref{ene}.

Furthermore, the dominance of this saddle would give the refined even R\'enyi negativity~\er{ste} as
\be\la{st2ne}
S^{T_B(2n,\text{even,nsym})}_{AB} = n^2 \pa_n I\(M_2^{AB}, \g_{A_1}^{(n)}, \g_{B_2}^{(n)}\) = \fr{|\g_{A_1}^{(n)} \cup \g_{B_2}^{(n)}: M_2^{AB}|}{4G},
\ee
where $|\g_{A_1}^{(n)} \cup \g_{B_2}^{(n)}: M_2^{AB}|$ denotes the area of the conical defect $\g_{A_1}^{(n)} \cup \g_{B_2}^{(n)}$ with opening angle~$\frac{2\pi}n$ in the bulk solution with asymptotic boundary $M_2^{AB}$.
Taking the $n\to 1$ limit, we find the refined R\'enyi-2 negativity defined by \cref{st2}:
\be\la{st2e}
S^{T_B(2,\text{nsym})}_{AB} = \fr{|\g_{A_1}^\ext \cup \g_{B_2}^\ext: M_2^{AB}|}{4G} = S_A(\r_{AB}^2) + S_B(\r_{AB}^2),
\ee
where $|\g_{A_1}^\ext \cup \g_{B_2}^\ext: M_2^{AB}|$ is the sum of the areas of two extremal surfaces $\g_{A_1}$, $\g_{B_2}$ in the smooth bulk solution $\cB(M_2^{AB})$ with asymptotic boundary $M_2^{AB}$ (and no conical defect),%
\footnote{Under the standard assumption that $\cB(M_2^{AB})$ respects the $\bZ_2$ replica symmetry, in \cref{st2e} we may replace the extremal surface $\g_{B_2}$ (homologous to region $B$ on the second copy in $M_2^{AB}$) by $\g_{B,1}$ (homologous to region $B$ on the first copy), as they have the same area.}
which is the same as the bulk geometry for calculating the second R\'enyi entropy on $AB$.  In the last equality, these two areas are related by the RT formula to $S_A(\r_{AB}^2)$ and $S_B(\r_{AB}^2)$, the von Neumann entropy of $A$ and $B$, respectively, in the ``doubled'' state $\r_{AB}^2 / \tr (\r_{AB}^2)$.  Intuitively, the RT formula applies here because $\cB(M_2^{AB})$ is the semiclassical geometry dual to the doubled state; more precisely, the applicability of the RT formula here may be derived by the replica method.

We expect that this $\bZ_n$ symmetric saddle contributes dominantly to the even R\'enyi negativity~\er{neven} deep in the case where the mutual information between $A$ and $B$ is positive, while the full replica symmetric one studied in \cref{ssecsym} dominates deep in the zero mutual information case.
However, the exact point of the phase transition generally depends on $k$ and is likely difficult to solve explicitly due to nontrivial gravitational backreaction.

\medskip

We can again find tractable solutions in fixed-area states where the areas of all three relevant codimension-2 surfaces $\g_A$, $\g_B$, and $\g_{AB}$ are fixed to particular values, which we call $|\g_A|$, $|\g_B|$, and $|\g_{AB}|$, respectively.
In such states, the on-shell action 
is simply
\begin{align*}
I\(M_2^{AB}, \g_{A_1}^{(n)}, \g_{B_2}^{(n)}\) = 2 I(M_1) + \fr{|\g_{AB}|}{4G} + \(1- \fr{1}{n}\) \fr{|\g_A|+|\g_B|}{4G},
\end{align*}
and the even R\'enyi negativity~\er{nke} simplifies to
\be\la{nefa}
\log N_{2n}^{(\text{even,nsym})} = -n \frac{|\g_{AB}|}{4G} -(n-1) \frac{|\g_A|+|\g_B|}{4G},
\ee
which again matches with the RTN result in the even case~\er{eq:logNeven} in the positive mutual information case under the identification $\log D=\frac1{4G}$.
Comparing \cref{nefa} with the replica symmetric contribution \er{nsfa}, we find that the $\bZ_n$ symmetric saddle considered here dominates%
\footnote{This is manifestly true when $n$ is a positive integer.
One might worry whether a phase transition could occur within $0<n<1$ and affect the logarithmic negativity~\er{enfa} (which is obtained as $n\to 1/2$).
The following argument shows that this cannot occur.  The fact that the R\'enyi entropy $S^{(k)}$ is monotonically non-increasing with $n$, applied to the density matrix $(\r_{AB}^{T_B})^2 / N_2$, implies $3\log N_2 -\log N_4 \leq 2E_N \leq \log N_0 +\log N_2$.  Using the values of $N_2$, $N_4$ from \cref{nefa} as well as $N_0 = \rk \rho_{AB}^{T_B} \leq \rk \rho_A \rk \rho_B = \exp(S^{(0)}_A +S^{(0)}_B) = \exp\[(|\g_A|+|\g_B|)/(4G)\]$, we find that $E_N$ is bounded by the right-hand side of \cref{enfa} from both above and below (at the leading order in $G$).  This proves \cref{enfa}.} if the mutual information
\begin{align*}
  I(A:B) = \frac{|\g_A| + |\g_B| - |\g_{AB}|}{4G}
\end{align*}
is nonzero at order~$O(1/G)$, whereas the full replica symmetric saddle studied in the previous subsection dominates if $I(A:B)$ vanishes at order $O(1/G)$.
Taking the $n\to\frac12$ limit in \cref{nefa}, we obtain the logarithmic negativity~\er{ene} in fixed-area states as simply
\be\la{enfa}
E_N^\nsym = \fr{|\g_A|+|\g_B| - |\g_{AB}|}{8G} = \fr{1}{2} I(A:B),
\ee
which again agrees with the RTN result in \cref{eq:negativityasymptotics}.
Furthermore, the refined even R\'enyi negativity~\er{st2ne} becomes independent of~$n$:
\be
S^{T_B(2n,\text{even,nsym})}_{AB} = \fr{|\g_A|+|\g_B|}{4G} = S_A + S_B,
\ee
which is also the value of the refined R\'enyi-2 negativity~\er{st2e}:
\be\la{st2fa}
S^{T_B(2,\text{nsym})}_{AB} = \fr{|\g_A|+|\g_B|}{4G} = S_A + S_B.
\ee
In fact, as we may see from \cref{st2e}, \cref{st2fa} holds in a much larger class of fixed-area states, where only the area of one codimension-2 surface $\g_{AB}$ is fixed.

\subsection{Replica nonsymmetric saddle for odd \texorpdfstring{$k$}{k}}\label{sec:hodd}
In this subsection, we focus on the case of odd $k=2n-1$.
To obtain a replica nonsymmetric saddle with asymptotic boundary $M_{2n-1}^{A,B}$, we use the following cutting and gluing procedure.

\begin{figure}[t]
\centering
\includegraphics[width=0.9\textwidth]{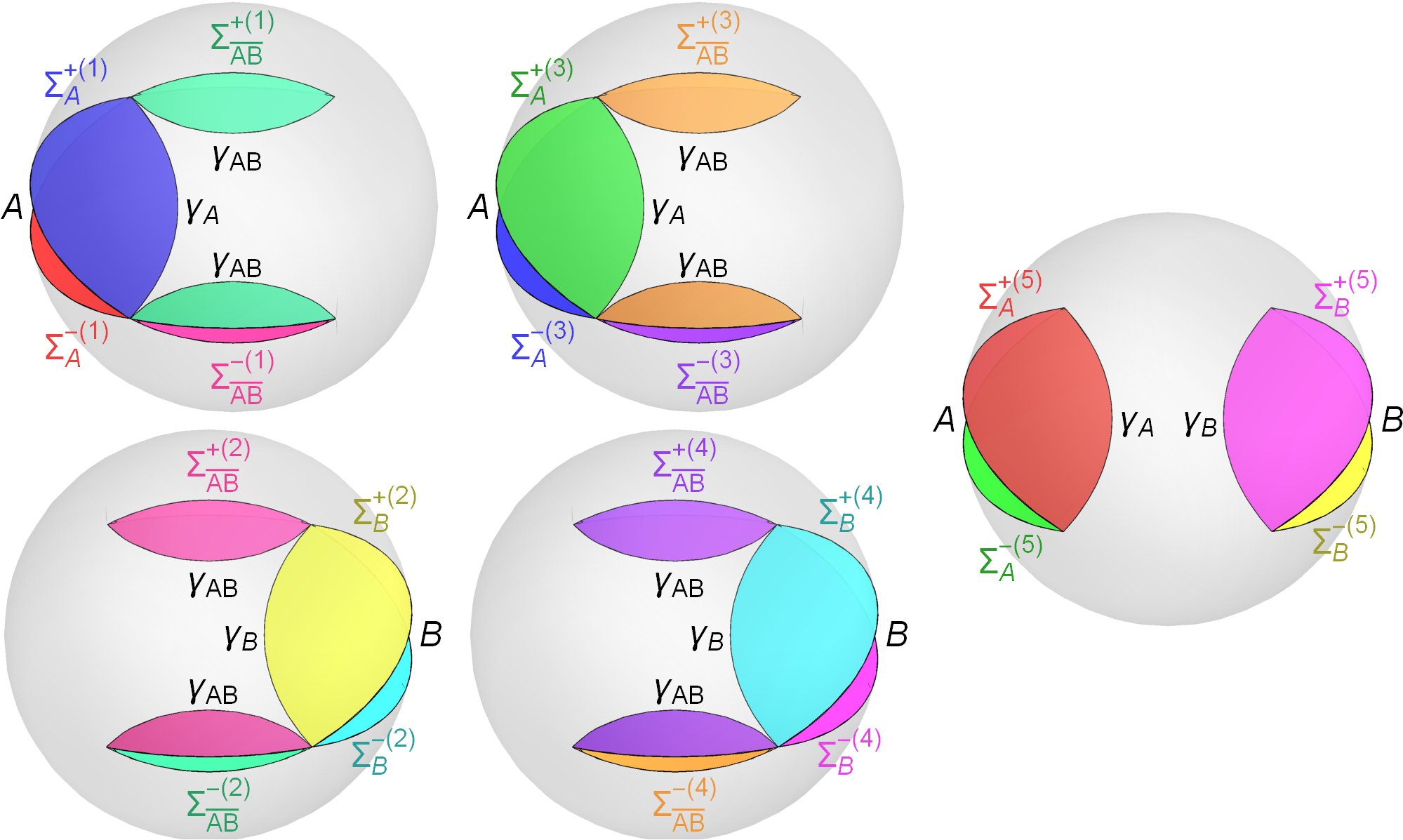}
\caption{The topology of a replica nonsymmetric saddle contributing to holographic R\'enyi negativity $N_k$ for $k=5$.
The 5 copies are first cut along $\S_A$, $\S_B$, $\S_{\ol{AB}}$, and then the cut surfaces $\S_A^{\pm(i)}$, $\S_B^{\pm(i)}$, and $\S_{\ol{AB}}^{\pm(i)}$ with the same color are glued together (cyclically along $\S_A$ in copies $1$, $3$, $5$, anti-cyclically along $\S_B$ in copies $2$, $4$, $5$, and pairwise along $\S_{\ol{AB}}$ between copies $2j-1$ and $2j$ for $j=1,2$).
Note that the surface $\S_{\ol{AB}}$ shown here has two connected components on each copy.}
\label{figodd}
\end{figure}

Starting with $2n-1$ copies of the original bulk spacetime $\cB_1$, we again cut each of them along three non-overlapping codimension-1 surfaces $\S_A$, $\S_B$, and $\S_{\ol{AB}}$ (defined as in the previous subsection).
We will focus on a particularly simple gluing procedure as follows (\cref{figodd}).
Let us label the copies by $1,2,\dots,2n-1$.
Along $\S_A$, we glue the $n$ odd copies labeled by $1,3,\dots,2n-1$ cyclically, and we glue each of the remaining copies to itself.
Along $\S_B$, we glue the $n-1$ copies labeled by $2,4,\dots,2n-2$ as well as $2n-1$ anti-cyclically, and we glue each of the remaining copies to itself.
Along $\S_{\ol{AB}}$, we glue the first two copies, then the next two copies, and so on until we reach the last copy, which we glue to itself.
It is manifest that this gluing procedure breaks the $\bZ_{2n-1}$ replica symmetry completely.

Again we may verify that the cutting and gluing procedure described above indeed leads to a bulk manifold with asymptotic boundary $M_{2n-1}^{A,B}$.
To see this, we note that the boundary entangling surface $\pa A$ is asymptotically approached by both $\g_{A}$ and $\g_{AB}$, and therefore the boundary permutation around $\pa A$ (which determines the gluing procedure along $A$ on the boundary) is the composition of the permutation $\tau = (12)(34)\cd(2n\!-\!3 ~ 2n\!-\!2)$ around~$\g_{AB}$ followed by the permutation $X \tau = (1~3~\dots~2n\!-\!1)$ around~$\g_A$.
This composition is the full $(2n-1)$-cycle $X = (1~2~\dots~2n\!-\!1)$, precisely what is needed to glue all $2n-1$ boundary copies cyclically along $A$.
Similarly, the boundary permutation around $\pa B$ is the composition of the permutation $(12)(34)\cd(2n-3,2n-2)(2n-1)$ around $\g_{AB}$ followed by the permutation $(2n-1,2n-2,2n-4,\cd,2)(1)(3)\cd(2n-3)$ around $\g_B$.
This composition is $(2n-1,2n-2,\cd,1)$, correctly gluing all $2n-1$ boundary copies anti-cyclically along $B$.
Again we see that even though the gluing procedure in the bulk manifestly breaks the $\bZ_{2n-1}$ replica symmetry, it is preserved on the boundary.

As before, we impose the equations of motions after fixing the bulk topology using the cutting and gluing procedure.
We will call this replica nonsymmetric saddle $\cB_{2n-1}^\nsym$.
As mentioned above, this saddle breaks the $\bZ_{2n-1}$ replica symmetry completely and does not preserve any nontrivial subgroup.
Therefore, there is no quotient space that we may use in this case to provide an analytic continuation away from integer $n$.
Thus we will not provide a more explicit expression for the contribution of this replica nonsymmetric saddle to the odd R\'enyi negativity~\er{nkr} in general cases than the following:
\be\la{nko}
  \log N_{2n-1}^{(\text{odd,nsym})} = -I\[\cB_{2n-1}^\nsym\] +(2n-1) I\[\cB_1\].
\ee

Nonetheless, we can solve this in fixed-area states where the areas of $\g_A$, $\g_B$, and $\g_{AB}$ are fixed to values $|\g_A|$, $|\g_B|$, and $|\g_{AB}|$, respectively.
In such states, the on-shell action of the replica nonsymmetric saddle $\cB_{2n-1}^\nsym$ is simply
\begin{align*}
I\[\cB_{2n-1}^\nsym\] = (2n-1) I\[\cB_1\] + (n-1) \fr{|\g_A|+|\g_B|+|\g_{AB}|}{4G}.
\end{align*}
Hence the odd R\'enyi negativity~\er{nko} simplifies to
\be\la{nofa}
\log N_{2n-1}^{(\text{odd,nsym})} = -(n-1) \frac{|\g_A|+|\g_B|+|\g_{AB}|}{4G},
\ee
which again matches with the RTN result in the odd case~\er{eq:logNodd} in the positive mutual information case under the identification $\log D=\frac1{4G}$.
Comparing \cref{nofa} with the replica symmetric contribution \er{nsfa}, we find that the replica nonsymmetric saddle considered here dominates if the mutual information $I(A:B)$ is nonzero at order $O(1/G)$, whereas the replica symmetric saddle studied in \cref{ssecsym} dominates if $I(A:B)$ vanishes at order $O(1/G)$.

Using \cref{nofa}, we find that the refined odd R\'enyi negativity~\er{sto} in fixed-area states becomes independent of~$n$:
\begin{align*}
S^{T_B(2n-1,\text{odd,nsym})}_{AB} = \fr{|\g_A|+|\g_B|+|\g_{AB}|}{8G} = \fr{1}{2}(S_A+S_B+S_{AB}),
\end{align*}
which is then also the value of the partially transposed entropy defined by~\er{st}:
\be\la{stfa}
S^{T_B\nsym}_{AB} = \fr{|\g_A|+|\g_B|+|\g_{AB}|}{8G} = \fr{1}{2}(S_A+S_B+S_{AB}).
\ee

We would like to conjecture that \cref{stfa} holds even in general semiclassical states where areas are not fixed.
The intuitive reason is that even though the replica nonsymmetric saddle $\cB_{2n-1}^\nsym$ is complicated to solve due to strong gravitational backreaction, taking the limit of $n\to 1$ in some appropriate way we should find that the backreaction become weak and the odd R\'enyi negativity~\er{nko} could be governed by the areas of the extremal surfaces $\g_A$, $\g_B$, and $\g_{AB}$ in \cref{stfa} to linear order in $n-1$.

We can show \cref{stfa} for general (non-fixed-area) states under the assumption of a certain ``diagonal approximation'' similar to the one used in~\cite{Marolf:2020vsi,Dong:2020iod}.
To do this, we write a general state $|\y\>$ as a superposition of fixed-area states $|\ol{\g_A},\ol{\g_B},\ol{\g_{AB}}\>$, where the areas of $\g_A$, $\g_B$, and $\g_{AB}$ are fixed to values $\ol{\g_A}$, $\ol{\g_B}$, and $\ol{\g_{AB}}$, respectively:%
\footnote{We previously called these values $|\g_A|$, $|\g_B|$, and $|\g_{AB}|$, but we have now changed their names as it would be awkward to write fixed-area states as $\big||\g_A|,|\g_B|,|\g_{AB}|\big\>$.}
\be\la{ysuper}
|\y\> = \sum_{\ol{\g_A},\ol{\g_B},\ol{\g_{AB}}} \sqrt{P(\ol{\g_A},\ol{\g_B},\ol{\g_{AB}})} |\ol{\g_A},\ol{\g_B},\ol{\g_{AB}}\>.
\ee
Here the probability $P(\ol{\g_A},\ol{\g_B},\ol{\g_{AB}})$ can be determined by a gravitational path integral with the three areas fixed:
\begin{align*}
P(\ol{\g_A},\ol{\g_B},\ol{\g_{AB}}) = \fr{1}{Z} \int_{\ol{\g_A},\ol{\g_B},\ol{\g_{AB}}} Dg \, e^{-I[g]},
\end{align*}
where $Z$ is calculated by the same integral but with the three areas unfixed.
Now we trace out $\ol{AB}$ and form the partial transpose of the density operator $\r_{AB}$:
\begin{equation}\la{rabts}
\begin{aligned}
\r_{AB}^{T_B} = \sum_{\ol{\g_A},\ol{\g_B},\ol{\g_{AB}},\ol{\g_A}',\ol{\g_B}',\ol{\g_{AB}}'} &\sqrt{P(\ol{\g_A},\ol{\g_B},\ol{\g_{AB}})P(\ol{\g_A}',\ol{\g_B}',\ol{\g_{AB}}')}\\
\times &\(\tr_{\ol{AB}} |\ol{\g_A},\ol{\g_B},\ol{\g_{AB}}\> \<\ol{\g_A}',\ol{\g_B}',\ol{\g_{AB}}'|\)^{T_B}.
\end{aligned}
\end{equation}
We may further separate this sum into diagonal contributions with $(\ol{\g_A},\ol{\g_B},\ol{\g_{AB}})=(\ol{\g_A}',\ol{\g_B}',\ol{\g_{AB}}')$ and the remaining, off-diagonal contributions:%
\footnote{As $\ol{\g_{AB}}$ can be reconstructed from $\ol{AB}$, the sum in \cref{rabts} must be diagonal in $\ol{\g_{AB}}$, and the only off-diagonal contributions come from $\ol{\g_A}\neq \ol{\g_A}'$ or $\ol{\g_B}\neq \ol{\g_B}'$.}
\be\la{rabtf}
\r_{AB}^{T_B} = \sum_{\ol{\g_A},\ol{\g_B},\ol{\g_{AB}}} P(\ol{\g_A},\ol{\g_B},\ol{\g_{AB}}) \, \r_{AB}^{T_B}(\ol{\g_A},\ol{\g_B},\ol{\g_{AB}}) + \text{(off-diagonal terms)},
\ee
where $\r_{AB}^{T_B}(\ol{\g_A},\ol{\g_B},\ol{\g_{AB}})$ is the partially transposed density operator of the fixed-area state $|\ol{\g_A},\ol{\g_B},\ol{\g_{AB}}\>$.
Now we assume that, for the purpose of calculating the partially transposed entropy $S^{T_B}$ in the state $\r_{AB}$ to leading order in~$G$, we may neglect the off-diagonal terms in \cref{rabtf} and approximate
\be\la{rabtd}
\r_{AB}^{T_B} \ap \r_{AB,\text{diag}}^{T_B} \eq \sum_{\ol{\g_A},\ol{\g_B},\ol{\g_{AB}}} P(\ol{\g_A},\ol{\g_B},\ol{\g_{AB}}) \, \r_{AB}^{T_B}(\ol{\g_A},\ol{\g_B},\ol{\g_{AB}}).
\ee
We will refer to this as the \emph{diagonal approximation}.
The sum in \cref{rabtd} is actually a direct sum since $\r_{AB}^{T_B}(\ol{\g_A},\ol{\g_B},\ol{\g_{AB}}) \r_{AB}^{T_B}(\ol{\g_A}',\ol{\g_B}',\ol{\g_{AB}}')=0$ if $\ol{\g_A} \neq \ol{\g_A}'$, $\ol{\g_B} \neq \ol{\g_B}'$, or $\ol{\g_{AB}} \neq \ol{\g_{AB}}'$, because the areas $\ol{\g_A},\ol{\g_B},\ol{\g_{AB}}$ all correspond to Hermitian operators that can be reconstructed on $AB$ according to the entanglement wedge reconstruction theorem~\cite{dong2016reconstruction}.
Therefore, under the diagonal approximation the partially transposed entropy becomes
\begin{equation}\la{std}
\begin{aligned}
S^{T_B\nsym}_{AB} &= \sum_{\ol{\g_A},\ol{\g_B},\ol{\g_{AB}}} P(\ol{\g_A},\ol{\g_B},\ol{\g_{AB}}) S^{T_B\nsym}_{AB}(\ol{\g_A},\ol{\g_B},\ol{\g_{AB}}) \\
&- \sum_{\ol{\g_A},\ol{\g_B},\ol{\g_{AB}}} P(\ol{\g_A},\ol{\g_B},\ol{\g_{AB}}) \log P(\ol{\g_A},\ol{\g_B},\ol{\g_{AB}}),
\end{aligned}
\end{equation}
where the second term -- the entropy of the probability distribution $P(\ol{\g_A},\ol{\g_B},\ol{\g_{AB}})$ -- is at most of order $\log(G)$ if we allow in each fixed-area state an area window that is polynomially small in~$G$, and thus the number of terms in the sum~\er{std} is polynomial in $\frac1G$.
If we accordingly neglect the last term in \cref{std} and use \cref{stfa} for fixed-area states, we obtain
\begin{align*}
S^{T_B\nsym}_{AB} = \sum_{\ol{\g_A},\ol{\g_B},\ol{\g_{AB}}} P(\ol{\g_A},\ol{\g_B},\ol{\g_{AB}}) \fr{\ol{\g_A}+\ol{\g_B}+\ol{\g_{AB}}}{8G}.
\end{align*}
This confirms \cref{stfa} under the diagonal approximation once we recognize that the areas $|\g_A|$, $|\g_B|$, and $|\g_{AB}|$ of the RT surfaces can be computed as the averages of $\ol{\g_A}$, $\ol{\g_B}$, and $\ol{\g_{AB}}$ under the probability distribution~$P(\ol{\g_A},\ol{\g_B},\ol{\g_{AB}})$.

Another piece of evidence for the conjecture~\er{stfa} to hold in non-fixed-area states is provided by the case where $A$ and $B$ are adjacent intervals in the vacuum state of a 2-dimensional CFT.
In this case, the result is completely fixed by the conformal symmetry~\cite{Calabrese:2012ew} and agrees with \cref{stfa}.

\section{Discussion}\label{sec:discussion}
In this paper we studied entanglement negativity and its R\'enyi generalizations in holography for general subregions $A$ and $B$.
We found that these quantities are nontrivial (and large) in the phase where the entanglement wedge of $AB$ connects~$A$ with~$B$, corresponding to a positive mutual information between~$A$ and~$B$ according to the Ryu-Takayanagi formula.

In random tensor networks as toy models of holography, we found in this phase simple expressions for the odd and even R\'enyi negativities in the large bond dimension limit (\cref{eq:logNodd,eq:logNeven}):
\begin{align}
    \log N_{2n-1}^\odd &\simeq-(n-1)\log(D)\left(\left|\gamma_{AB}\right|+\left|\gamma_A\right|+\left|\gamma_B\right|\right)
    ~+~\log a_{2n-1},\label{eq:logNodd2}\\
    \log N_{2n}^\even &\simeq-n\log(D)\left|\gamma_{AB}\right|-(n-1)\log(D)\left(\left|\gamma_A\right|+\left|\gamma_B\right|\right)
    ~+~\log a_{2n}.\label{eq:logNeven2}
\end{align}
This leads to a simple logarithmic negativity (\cref{eq:negativityasymptotics}):
\begin{align}\label{eq:negativityasymptotics2}
    E_N(\rho_{AB}) \simeq \frac12I(A:B) + \log \frac{8}{3\pi}.
\end{align}
We also showed that, unlike the entanglement spectrum, the eigenvalue distribution of the partial transpose~$\rho_{AB}^{T_B}$ does \emph{not} become flat in the large bond dimension limit but rather follows a Wigner semicircle law; see the discussion surrounding \cref{eq:empirical eigs}.
This allowed us to give a rigorous derivation of \cref{eq:negativityasymptotics2} that did not rely on analytic continuation.

In theories holographically dual to quantum gravity, we found that replica nonsymmetric saddle points play a dominant role in determining R\'enyi negativities in the nontrivial phase where the mutual information between~$A$ and~$B$ is positive.
We expressed the odd and even holographic R\'enyi negativities (as well as their special limits such as logarithmic negativity) in terms of on-shell actions of specific replica nonsymmetric saddles in \cref{nke,nko}.
These saddles are generally difficult to solve explicitly due to gravitational backreaction.
However, in fixed-area states they can be solved easily, and we found results for the odd and even R\'enyi negativities (including special limits such as logarithmic negativity) that coincide with the formulas \cref{eq:logNodd2,eq:logNeven2,eq:negativityasymptotics2} for random tensor networks.

Furthermore, we identified a number of tractable limits of R\'enyi negativities in general semiclassical states of holographic theories.
Even though the relevant saddle point is difficult to solve because the areas of RT surfaces are not fixed in these general states, we were able to use a residual replica symmetry to analytically continue its on-shell action in the replica number at least for the even R\'enyi negativities.
This leads to relatively simple expressions \cref{ene,st2e} for the logarithmic negativity and refined R\'enyi-2 negativity, in terms of a doubled bulk saddle relevant for calculating the second R\'enyi entropy of $AB$ (with possible insertions of additional cosmic branes).
Moreover, the odd R\'enyi negativities have a special limit which we called the partially transposed entropy, and we found a simple expression \cref{stfa} for it under a diagonal approximation.

So far we have focused deep within different phases of negativity (such as the replica symmetry breaking phase).  It would be very interesting to analyze the behavior of negativity near phase transitions using ideas from~\cite{Penington:2019kki,Marolf:2020vsi,Dong:2020iod}; we leave this to future work.

\subsection{Comparison with previous results}

As mentioned earlier, several important prior works have explored the negativity in conformal field theories and in holography.
We now give a detailed comparison between our results with those of prior works.

\subsubsection{Pure states}

It is a well-known result~\cite{vidal2002computable} that for a pure state on $AB$, the logarithmic negativity is simply $S_{1/2}$, the R\'enyi entropy of order~$1/2$ for either subsystem~$A$ or~$B$.
Our holographic results easily reproduce this special situation.
In the nontrivial case when $A$ and $B$ have positive mutual information (equivalently, when $A$ and $B$ are connected through the bulk), the logarithmic negativity as given by \cref{ene} becomes
\begin{align*}
E_N = I(M_1) - \fr{1}{2} I\(M_2^{AB}, \g_{A_1}^{(\fr{1}{2})}, \g_{B_2}^{(\fr{1}{2})}\) = I(M_1) - I\(M_1,\g_{A}^{(\fr{1}{2})}\),
\end{align*}
where the second equality comes from the following argument.
As $AB$ is in a pure state, the $2$-fold cover $M_2^{AB}$ branched over $AB$ is actually a disjoint union of two copies of the original boundary manifold $M_1$.
Thus the corresponding bulk solution with conical defects on $\g_{A_1}$ (homologous to $A$ in the first copy) and $\g_{B_1}$ (homologous to $B$ in the second copy) decomposes into two disjoint bulk components, one for each boundary copy, with $\g_{A_1}$ in the first bulk component and $\g_{B_1}$ in the second.
These two bulk components are in fact identical, as being homologous to $A$ is equivalent to being homologous to $B$ in a pure state.
The total on-shell action is therefore twice the action of a single bulk component -- which we write as $I(M_1,\g_{A}^{(\fr{1}{2})})$ because the component is characterized as the solution with boundary $M_1$ and a conical defect on $\g_{A}^{(\fr{1}{2})}$ with opening angle $4\pi$.
This is precisely the same bulk saddle that appears in the holographic calculation of R\'enyi entropy~\cite{Lewkowycz:2013nqa,dong2016gravity} of order $1/2$, and we find
\begin{align*}
E_N = I(M_1) - I\(M_1,\g_{A}^{(\fr{1}{2})}\) = S_{1/2}.
\end{align*}

\subsubsection{Adjacent intervals in two dimensions}

When $A$ and $B$ are two adjacent intervals in the vacuum state of a 2-dimensional CFT, the negativity and its R\'enyi generalizations are determined by universal 3-point functions of twist operators~\cite{Calabrese:2012ew}.
In particular, the even R\'enyi negativity~\er{neven} in such cases is
\be\la{neai}
N^{\even}_{k} (\r_{AB}) \propto (\ell_A \ell_B)^{-\fr{c}{6}\(\fr{k}{2}-\fr{2}{k}\)} (\ell_A+\ell_B)^{-\fr{c}{6} \(\fr{k}{2}+\fr{1}{k}\)},
\ee
where $\ell_A$ and $\ell_B$ denote the length of $A$ and $B$, respectively.
The logarithmic negativity~\er{en} obtained by taking the $k\to1$ limit is
\be\la{enai}
E_N = \lim_{k\to1} \log N^{\even}_{k} = \fr{c}{4} \log \(\fr{\ell_A \ell_B}{\ell_A+\ell_B}\) + \text{const},
\ee
and the refined R\'enyi-2 negativity~\er{st2} obtained by taking the $k\to2$ limit is
\begin{align*}
S^{T_B(2)}_{AB} = -\lim_{k\to2} k^2 \pa_k \(\fr{1}{k} \log N_{k}^\even\) = \fr{c}{6} \log\(\fr{\ell_A^2 \ell_B^2}{\ell_A+\ell_B}\) +\text{const}.
\end{align*}
Furthermore, the odd R\'enyi negativity~\er{nodd} is
\be\la{noai}
N^{\odd}_{k} (\r_{AB}) \propto \[\ell_A \ell_B (\ell_A+\ell_B)\]^{-\fr{c}{12}\(k-\fr{1}{k}\)}.
\ee
The partially transposed entropy~\er{st} obtained by taking the $k\to1$ limit is
\be\la{stai}
S^{T_B}_{AB} = -\lim_{k\to1} \pa_k \log N^{\odd}_{k} = \fr{c}{6} \log \[\ell_A \ell_B (\ell_A+\ell_B)\] +\text{const}.
\ee

All of these are reproduced precisely by our holographic results~\er{nke} and~\er{nko}.
This can be verified by direct calculations, or by noting that our holographic results automatically satisfy the conformal symmetry which completely fixes the R\'enyi negativities~\er{neai} and~\er{noai} for adjacent intervals.
Additionally, it is worth noting that the partially transposed entropy~\er{stai} agrees with our holographic conjecture that \cref{stfa} holds for general non-fixed-area states (which we showed under the diagonal approximation assumption).

\subsubsection{Adjacent regions in general dimensions}

The result~\er{enai} for the logarithmic negativity was conjectured in~\cite{Jain:2017aqk,Jain:2017xsu} to generalize to $E_N = \fr{3}{4} I(A:B)$ for adjacent regions $A$, $B$ in holographic theories with general dimensions, in a general state (not necessarily the vacuum).
This conjecture can be thought of as a linear combination of the areas of the RT surfaces for $A$, $B$, and $AB$.

Our holographic result~\er{ene} precisely determines the logarithmic negativity in all dimensions, but it generally involves nontrivial gravitational backreaction and does not simplify into a linear combination of RT surface areas.
A notable exception is the case of fixed-area states, for which our result simply becomes \cref{enfa}:
\begin{align*}
E_N = \fr{|\g_A|+|\g_B| - |\g_{AB}|}{8G} = \fr{1}{2} I(A:B).
\end{align*}
Therefore in this special case, our prefactor $1/2$ disagrees with the prefactor $3/4$ in the conjecture of~\cite{Jain:2017aqk,Jain:2017xsu}.

\subsubsection{Disjoint regions in general dimensions}

A recent conjecture for the logarithmic negativity was made in~\cite{Kudler-Flam:2018qjo,Kusuki:2019zsp}, which can be applied to the general case of disjoint regions $A$, $B$.
The conjecture states that in holographic theories, the logarithmic negativity is given by a backreacting entanglement wedge cross section, or more precisely by the R\'enyi-$1/2$ reflected entropy:
\be\la{enconj}
E_N^{\text{conj}} = \frac12 S_R^{(1/2)}.
\ee
Here we recall that any state $\r_{AB}$ admits a canonical purification $\ket{\sqrt{\r_{AB}}}$ in the Hilbert space $\cH_A \otimes \cH_{A^*} \otimes \cH_B \otimes \cH_{B^*}$, and the R\'enyi reflected entropy~\cite{dutta2019canonical} is defined as the R\'enyi entropy of $AA^*$ in this state:
\begin{align*}
S_R^{(k)} = S^{(k)}_{AA^*}\(|\sqrt{\r_{AB}}\>\).
\end{align*}

Focusing on the nontrivial phase where the mutual information between $A$ and $B$ is positive, our holographic result~\er{ene} for logarithmic negativity is different from the conjecture of~\cite{Kudler-Flam:2018qjo,Kusuki:2019zsp,kudler2020entanglement}.
The difference is not only in the relevant bulk saddles, but also in the predicted values for logarithmic negativity.
This is most obvious in fixed-area states where the areas of the RT surfaces for $A$, $B$, and $AB$, as well as the area of the entanglement wedge cross section, have been fixed to particular values.
In these fixed-area states, our result~\er{ene} for logarithmic negativity reduces to \cref{enfa}:
\be\la{enfa2}
E_N = \fr{|\g_A|+|\g_B| - |\g_{AB}|}{8G} = \fr{1}{2} I(A:B),
\ee
whereas the right-hand side of the conjecture~\er{enconj} simply becomes $E_W$, defined as the area of the entanglement wedge cross section in the original geometry (which is one of the areas being fixed) divided by $4G$:
\be\la{enconjfa}
E_N^{\text{conj}} = E_W.
\ee
These two answers obviously disagree.  In particular, as we vary the distance between~$A$ and~$B$ and go through a phase transition into the trivial phase where $I(A:B)=0$ at the leading order in $G$, our predicted logarithmic negativity~\er{enfa2} is continuous, whereas the conjectured dual~\er{enconjfa} in terms of the entanglement wedge cross section experiences a discontinuity.
Such a discontinuity may violate a continuity bound for logarithmic negativity proved in~\cite{Lu:2020jza}; we leave a careful analysis of this to future work.\footnote{We note that in a revision of Ref.~\cite{kudler2020entanglement} the authors have discussed this problem. We thank Shinsei Ryu for sharing that update with us.}

A different, but analogous, conjecture was made in~\cite{Tamaoka:2018ned} for the partially transposed entropy (which was called the ``odd entanglement entropy'' in~\cite{Tamaoka:2018ned}).  The conjecture states that in holographic theories, the difference between the partially transposed entropy and the von Neumann entropy of $AB$ is given by the area of the entanglement wedge cross section:
\be\la{stconj}
S^{T_B,\text{conj}}_{AB} - S_{AB} = E_W.
\ee
Our holographic result~\er{stfa}:
\be\la{stfa2}
S^{T_B}_{AB} = \fr{|\g_A|+|\g_B|+|\g_{AB}|}{8G} = \fr{1}{2}(S_A+S_B+S_{AB}),
\ee
which can be equivalently written as
\be
S^{T_B}_{AB} - S_{AB} = \fr{1}{2} I(A:B),
\ee
disagrees with the conjecture~\er{stconj}.  In particular, although \cref{stfa2} is strictly speaking a conjecture for general non-fixed-area states (which we needed to assume the diagonal approximation to prove), we have nonetheless presented a complete derivation of \cref{stfa2} for fixed-area states where the areas of the RT surfaces for $A$, $B$, and $AB$ are fixed to particular values.  It is clear that in these fixed-area states our result~\er{stfa2} disagrees with the conjecture~\er{stconj}.

For disjoint intervals in 2-dimensional holographic CFTs, a derivation of the conjecture~\er{enconj} was presented in~\cite{Kusuki:2019zsp}, and a derivation of the conjecture~\er{stconj} was given in~\cite{Tamaoka:2018ned}.
These derivations seem to correspond to bulk calculations using replica symmetric saddles.
Our results were derived from replica nonsymmetric saddles that give dominant contributions in the bulk calculation, so it is not surprising that our results disagree with the conjectures~\er{enconj} and~\er{stconj}.
It would be interesting to try to reconcile these answers by adding replica nonsymmetric contributions into the derivations of~\cite{Kusuki:2019zsp,Tamaoka:2018ned} in two dimensions.

\section*{Acknowledgments}
We would like to thank Tom Faulkner, Don Marolf, Sepehr Nezami, Pratik Rath for many interesting discussions.
MW would like to thank C\'ecilia Lancien for insights into the method of moments and geodesic permutations.
XD was supported in part by the National Science Foundation under Grant No.\ PHY-1820908 and by funds from the University of California. XLQ was supported by the National Science Foundation under Grant No.\ 1720504, and the Simons Foundation. MW acknowledges financial support by the NWO through Veni grant no.~680-47-459.
This work was developed in part at the Kavli Institute for Theoretical Physics which is supported in part by the National Science Foundation under Grant No.\ PHY-1748958.

\appendix
\section{Geodesics on the permutation group}\label{app:permutations}
In this section we first recall some known facts about geodesics on the permutation group~$S_k$ (see, e.g.,~\cite{nica2006lectures}).
Next, we characterize the set of permutations that are simultaneously on geodesics between the identity permutation, a $k$-cycle, and its inverse.
We will see in \cref{app:domains} that these permutations determine the ground state degeneracy of the spin model corresponding to the average of the negativity measure~$N_k$.

Given a permutation $g\in S_k$, define the \emph{length}~$\ell(g)$ as the minimal number of swaps which multiply to~$g$.
It is easy to see that
\begin{align}\label{eq:l plus chi}
  \ell(g)+\chi(g)=k
\end{align}
where $\chi(g)$ denotes the number of disjoint cycles in~$g$.
Then, $d(g,h) \equiv \ell(g^{-1}h)$ defines a \emph{metric} on~$S_k$.
For example,
\begin{align}\label{eq:perm dists}
  d(\id,X)=k-1, \quad
  d(\id,X^{-1})=k-1, \quad
  d(X,X^{-1})=\begin{cases}
    k-1, & \text{$k$ odd},\\
    k-2, & \text{$k$ even}.
  \end{cases}
\end{align}
where $\id$ denotes the identity permutation and $X=(1\,2\,\dots\,k)$ the standard $k$-cycle.

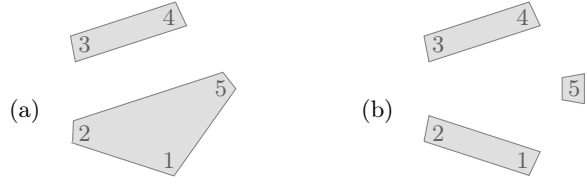
\begin{figure}
\centering
\raisebox{.8cm}{\footnotesize (a)}~~~\begin{tikzpicture}
  \foreach \x in {1,2,...,5} { 
  \node at (-\x*72:1cm) {\footnotesize\x}; }
  \draw[opacity=0.5,fill=lightgray] (71:1.22cm)--(145:1.22cm)--(159:1cm)--(57:1cm)--(71:1.22cm);
  \draw[opacity=0.5,fill=lightgray] (216:1.22cm)--(288:1.22cm)--(0:1.2cm)--(12:1.05cm)--(204:1.05cm)--(216:1.2cm);
\end{tikzpicture}
\qquad\qquad
\raisebox{.8cm}{\footnotesize (b)}~~~\begin{tikzpicture}
  \foreach \x in {1,2,...,5} { 
  \node at (-\x*72:1cm) {\footnotesize\x}; }
  \draw[opacity=0.5,fill=lightgray] (71:1.22cm)--(145:1.22cm)--(159:1cm)--(57:1cm)--(71:1.22cm);
  \draw[opacity=0.5,fill=lightgray] (71+144:1.22cm)--(145+144:1.22cm)--(159+144:1cm)--(57+144:1cm)--(71+144:1.22cm);
  \draw[opacity=0.5,fill=lightgray] (-10:1.15cm)--(+10:1.15cm)--(+10:0.85cm)--(-10:0.85cm)--(-10:1.15cm);
\end{tikzpicture}
\caption{\label{fig:NC} Partitions of $[k]=\{1,\dots,5\}$:
(a)~Non-crossing partition $P=\{\{1,2,5\},\{3,4\}\}$.
(b)~Non-crossing \emph{pairing} $P=\{\{1,2\},\{3,4\},\{5\}\}$.}
\end{figure}

We say that $(g_1,\dots,g_n)$ is a \emph{geodesic} if
\begin{align}\label{eq:geodesic}
  d(g_1,g_2) + \dots + d(g_{n-1},g_n) = d(g_1,g_n),
\end{align}
i.e., the triangle inequality is saturated.
The set of permutations~$g$ that are on a geodesic between $\id$ and $X$, i.e.,
\begin{align*}
  d(\id,g) + d(g,X) = k-1,
\end{align*}
are known to be in bijection with $\NC(k)$, the set of \emph{non-crossing partitions} of the set $[k] = \{1,\dots,k\}$.
See \cref{fig:NC},~(a) for an example of a non-crossing partition for~$k=5$.
The bijection is given as follows:
For any partition $P = \{P_1,\dots,P_n\}$, $[k] = \bigcup_{j=1}^n P_j$, define a corresponding permutation~$g_P$ as the product of disjoint cycles, one for each~$P_j$, permuting the elements of~$P_j$ in increasing order.
If $P$ is non-crossing then~$g_P$ is on a geodesic between~$\id$ and~$X$, and all permutations with this property can be obtained in this way.
E.g., if $P = \{\{1\}, \{2,5\}, \{3,4\}\}$ then $g_P = (2~5) (3~4)$.
From \cref{eq:l plus chi}, it is clear that $d(\id,P) = k - |P|$.
The number of non-crossing partitions is given by the \emph{Catalan numbers},
\begin{align}\label{eq:catalan}
  C_k \equiv |\NC(k)| = \frac 1 {k+1} \binom{2k}k.
\end{align}

In \cref{app:domains}, we will be interested in the permutations~$\tau$ that can simultaneously be on geodesics between $\id$, $X$, and $X^{-1}$.
That is,
\begin{equation}\label{eq:permtri}
\begin{aligned}
  d(\id,\tau) + d(\tau,X) &= k-1, \\
  d(\id,\tau) + d(\tau,X^{-1}) &= k-1, \\
  d(X,\tau) + d(\tau,X^{-1}) &= \begin{cases}
    k-1, & \text{$k$ odd},\\
    k-2, & \text{$k$ even},
  \end{cases}
\end{aligned}
\end{equation}
which is equivalent to
\begin{align}\label{eq:permtri eqv}
  d(\id,\tau) = \left\lfloor \frac k2 \right\rfloor, \quad
  d(X,\tau) = d(X^{-1},\tau) = \left\lceil \frac k2 \right\rceil - 1.
\end{align}
The first condition in \cref{eq:permtri} states that $\tau$ is geodesic, i.e., $\tau=g_P$ for some non-crossing partition $P\in\NC(n)$.
The second condition is equivalent to $d(\id,\tau^{-1}) + d(\tau^{-1},X) = k-1$, i.e., it states that $\tau^{-1}$ should also be on a geodesic between $\id$ and $X$.
We can invert $\tau=g_P$ cycle by cycle; the resulting cycles can be written in increasing order iff each set in the partition~$P$ is of length at most~2.
Now, the third condition in \cref{eq:permtri} if and only if $d(\id,\tau) = \left\lfloor \frac k2 \right\rfloor$.
But $d(\id,\tau)$ is simply the number of blocks of length~2.
We conclude that the permutations~$\tau$ that satisfy \cref{eq:permtri} (equivalently, \cref{eq:permtri eqv}) are precisely those that correspond to a non-crossing partition of~$[k]$ that contains only blocks of length two, except for a single block of length one if~$k$ is odd.
We call such a partition a \emph{non-crossing pairing}, the set of all non-crossing pairings by~$\NC_2(k)$, and its cardinality by $a_k$.
See \cref{fig:NC},~(b) for an example of a non-crossing pairing for~$k=5$.
If $k$ is even then there is a well-known bijection between $\NC_2(k)$ and $\NC(k/2)$.
If $k$ is odd then $a_k = k a_{k-1}$, since there are $k$ choices for the fixed point and then $a_{k-1}$ non-crossing pairings of the remaining symbols.
Thus:
\begin{align}\label{eq:num tau}
  a_k = \begin{cases}
  k \, C_{(k-1)/2}, & \text{$k$ odd},\\
  C_{k/2}, & \text{$k$ even}.
  \end{cases}
\end{align}
(E.g., for $k=3$ the only such permutations are the three swaps.)
Using \cref{eq:catalan} and expressing the binomial coefficient in terms of the beta function~$B(x,y)$, it is not hard to evaluate the analytic continuation of $k\to1$ of either expression:
\begin{align*}
  \lim_{k\to 1, \text{ odd}} a_k &= C_0 = \frac1{B(1,1)} = 1, \\
  \lim_{k\to 1, \text{ even}} a_k &= C_{1/2} = \frac83\frac1{B(1/2,1/2)} = \frac8{3\pi}.
\end{align*}

We end with a technical observation that we will use later when computing the absolute eigenvalue distribution of the partial transpose in \cref{app:spec}.
For permutations $g_A \in S_a$ and $g_B \in S_b$, write $g = (g_A,g_B)$ for the permutation that acts as $g_A$ on $\{1,\dots,a\}$ and as $g_B$ on $\{a+1,\dots,a+b\} \cong \{1,\dots,b\}$.
Then we claim that the permutations $\tau$ that are on a geodesic between $g = (g_A,g_B)$ and $h = (h_A,h_B)$ are exactly the permutations of the form $\tau = (\tau_A,\tau_B)$, where $\tau_A$ is on a geodesic between~$g_A$ and~$h_A$, and similarly for $\tau_B$.
In particular, there are $a_k^2$ many permutations that are simultaneously on geodesics between $\id$, $Y$, and $Y^{-1}$, where $Y=(X,X)$.

To prove the claim, we may assume without loss of generality that $g = \id$.
Now, if~$\tau$ is on a geodesic between~$\id$ and $h$, then each cycle of~$\tau$ is contained in a cycle of~$h$ (this follows, e.g., from \cite[Lemma 23.10]{nica2006lectures}).
In particular, $\tau$ is of the form $\tau = (\tau_A,\tau_B)$.
The claim now follows by observing that $d(\id, \tau) = d(\id, \tau_A) + d(\id, \tau_B)$ etc.

\section{Derivation of the classical spin model}\label{app:spin model}
In this section we derive the $S_k$-spin model described in the main text that calculates the average negativity for a random tensor network state~$\ket\Psi$.
The calculation is similar to~\cite{hayden2016holographic,nezami2016multipartite} but we spell out the details for the convenience of the reader.

Let $G=(V,E)$ be an undirected graph, where~$V$ is the vertex set and~$E$ the edge set.
We assume that $G$ is connected.
We are given a decomposition of the vertex set into bulk and boundary vertices, $V=V_b \cup V_\partial$.
We allow parallel edges in~$E$ but (for simplicity) no loops.
By slight abuse of notation, we write $e=\overline{xy}$ to indicate that~$e$ is an edge incident to vertices~$x$ and~$y$ (there can be more than one such edge).
For any subset of vertices~$\Gamma\subseteq V$, we write~$\partial\Gamma$ for the set of edges cut by~$\Gamma$ (i.e., one vertex is in~$\Gamma$ and the other in~$\Gamma^c=V\setminus\Gamma$).
To each vertex~$x$, we associate a Hilbert space~$\mathcal H_x = \bigotimes_{e \in \partial\{x\}} \mathcal H_{x,e}$, with one subsystem $\mathcal H_{x,e} = \mathbb C^D$ for each edge~$e$ incident to~$x$.
Let~$D_x$ denote the dimension of~$\mathcal H_x$ and define $D_b = \prod_{x\in V_b} D_x$.

We now define the random tensor network state~$\ket\Psi \in \mathcal H_\partial = \bigotimes_{x\in V_\partial} \mathcal H_x$.
For each bulk vertex~$x$, let $\ket{V_x}$ denote a Haar-random pure state in~$\mathcal H_x$.
For each edge~$e$, define a maximally entangled state~$\ket{L_e} = \frac1{\sqrt D}\sum_{i=1}^D \ket{ii} \in \mathcal H_{x,e} \ot \mathcal H_{y,e}$, where $x$ and $y$ denote the vertices of~$e$.
Then,
\begin{align}\label{eq:rtns}
  \ket\Psi \equiv D_b^{1/2} \Bigl( \otimes_{x\in V_b} \bra{V_x} \Bigr) \Bigl( \otimes_{e\in E} \ket{L_e} \Bigr),
\end{align}
where the normalization constant is chosen so that the state is normalized on average as~$D\to\infty$ (see below).

Fix an integer~$k\geq1$ and a partition of the boundary vertices into three subsets, $V_\partial = A\cup B\cup C$.
For each vertex $x\in V$ and permutation $g\in S_k$, we define $P_x(g)$ as the unitary operator that acts on~$\mathcal H_x^{\ot k}$ by permuting the $k$ tensor factors according to the permutation~$g$.
Note that $P_x(g) = \otimes_{e\in\partial\{x\}} P_{x,e}(g)$, where $P_{x,e}(g)$ denotes the corresponding permutation operator on~$\mathcal H_{x,e}^{\ot k}$.
We also write $P_\Gamma(g) = \otimes_{x\in\Gamma} P_x(g)$ for any~$\Gamma\subseteq V$.
We are interested in computing the average
\begin{align*}
  \overline{N'_k} \equiv \overline{\braket{\Psi^{\ot k}|P_A(X) \ot P_B(X^{-1}) \ot P_C(\id) |\Psi^{\ot k}}}
\end{align*}
where the prime is a reminder that $\ket\Psi$ is in general an \emph{unnormalized} quantum state on~$\mathcal H_\partial$.
As before, $\id$ denotes the identity permutation and $X=(1\,2\,\dots\,k)$ the standard~$k$-cycle.

To compute this average, we first recall that, by Schur's lemma,
\begin{align}\label{eq:haar}
  \overline{\ket{V_x}^{\ot k}\bra{V_x}^{\ot k}} = \frac1{D_{x,k}} \sum_{g\in S_k} P_x(g)
\end{align}
where $D_{x,k} \equiv \frac{(D_x+k-1)!}{(D_x-1)!}$ (note that $D_{x,1} = D_x$).
Then, using cyclicity of the trace and then \cref{eq:haar},
\begin{align}
\nonumber
  \overline{N'_k} &= \overline{\braket{\Psi^{\ot k}|P_A(X) \ot P_B(X^{-1}) \ot P_C(\id) |\Psi^{\ot k}}} \\
\nonumber
&= D_b^k \tr\left[\Bigl( \bigotimes_{e\in E} \ket{L_e}^{\ot k}\bra{L_e}^{\ot k} \Bigr) \Bigl(P_A(X) \ot P_B(X^{-1}) \ot P_C(\id) \ot \bigotimes_{x\in V_b} \overline{\ket{V_x}^{\ot k}\bra{V_x}^{\ot k}} \Bigr)\right]\\
\nonumber
&= \frac{D_b^k}{D_{b,k}} \sum_{\{g_x\}} \tr\left[\Bigl( \bigotimes_{e\in E} \ket{L_e}^{\ot k}\bra{L_e}^{\ot k} \Bigr) \Bigl(\bigotimes_{x\in V} P_x(g_x) \Bigr)\right]\\
\label{eq:same up to here}
&= \frac{D_b^k}{D_{b,k}} \sum_{\{g_x\}} \prod_{e=\overline{xy}\in E} \tr\left[\ket{L_e}^{\ot k}\bra{L_e}^{\ot k} \bigl( P_{x,e}(g_x) \ot P_{y,e}(g_y) \bigr)\right]\\
\nonumber
&= \frac{D_b^k}{D_{b,k}} \sum_{\{g_x\}} \prod_{e=\overline{xy}\in E} D^{-k+\chi(g_x g_y^{-1})}
= \frac{D_b^k}{D_{b,k}} \sum_{\{g_x\}} e^{-\log(D) \sum_{e=\overline{xy}\in E} d(g_x, g_y)},
\end{align}
where $D_{b,k} \equiv \prod_{x\in V_b} D_{x,k}$ (note that $D_{b,1} = D_b$), and the sum $\sum_{\{g_x\}}$ is over all configurations~$g_x\in S_k$ subject to the boundary conditions
\begin{align}\label{eq:bc}
  g_x = \begin{cases}
    X, &x \in A, \\
    X^{-1}, &x \in B, \\
    \id, &x \in C.
  \end{cases}
\end{align}
Thus,
\begin{align}\label{eq:spin model appendix}
  \braket{N'_k} = \frac{D_b^k}{D_{b,k}} Z_k, \quad
  Z_k \equiv \sum_{\{g_x\}} e^{-\log(D) E[\{g_x\}]}, \quad
  E[\{g_x\}] \equiv \sum_{e=\overline{xy}\in E} d(g_x, g_y).
\end{align}
Note that $D_{b,k} = D_b^k (1 + O(1/D))$, so $\braket{N'_k} = Z_k (1 + O(1/D))$ for large $D$.

For $A=B=\emptyset$ in particular, it is clear that $\overline{\braket{\Psi|\Psi}^k} = 1 + O(1/D)$ for all~$k$, so likewise~$\overline{\braket{\Psi|\Psi}^{2k}} = 1 + O(1/D)$.
It follows that $\braket{\Psi|\Psi}^k = 1 + o(1)$ with high probability as $D\to\infty$.
Thus, we have that $\overline{N_k} = Z_k + o(1)$, so we recover the spin model with action~\eqref{eq:spin model action}.

\section{Details of the domain wall calculation}\label{app:domains}
In this section, we will precisely determine the ground state energy and degeneracy of the spin model defined by \cref{eq:spin model appendix,eq:bc}.

Our main tool is a result from the theory of multicommodity flows.
Without loss of generality, assume that the degree of each vertex is even (if not, split each edge into two).
Then the result asserts that there exists a collection~$P=P_{AB}\cup P_{AC}\cup P_{BC}$ of $(\lvert\gamma_A\rvert+\lvert\gamma_B\rvert+\lvert\gamma_C\rvert)/2$ many \emph{edge-disjoint paths}, where $P_{AB}$ is a path between~$A$ and~$B$ and similarly for~$P_{AC}$ and~$P_{BC}$~\cite{kupershtokh1971generalization,lovasz1976some,cherkasski1977solution} (cf.~\cite{cui2018bit}).
Here, we write $\lvert\gamma_A\rvert$ for the size of a min-cut of $A$ (i.e., the number of edges crossing the cut), and similar for~$B$ and~$C$.

By the weak min-cut/max-flow duality, we always have that~$|P_{AB}| + |P_{AC}| \leq \lvert\gamma_A\rvert$, and similar for~$B$ and~$C$.
Since $|P| = (\lvert\gamma_A\rvert+\lvert\gamma_B\rvert+\lvert\gamma_C\rvert)/2$, it follows that these inequalities must be saturated:
\begin{align}\label{eq:entropies}
  |P_{AB}| + |P_{AC}| = \lvert\gamma_A\rvert, \quad
  |P_{AB}| + |P_{BC}| = \lvert\gamma_B\rvert, \quad
  |P_{AC}| + |P_{BC}| = \lvert\gamma_C\rvert.
\end{align}
This means that the paths in~$P$ simultaneously saturate min-cuts of $A$, $B$, and~$C$.
In turn:
\begin{align}\label{eq:path count}
  |P_{AB}| = \frac{\lvert\gamma_A\rvert + \lvert\gamma_B\rvert - \lvert\gamma_C\rvert}2,\quad
  |P_{AC}| = \frac{\lvert\gamma_A\rvert + \lvert\gamma_C\rvert - \lvert\gamma_B\rvert}2,\quad
  |P_{BC}| = \frac{\lvert\gamma_B\rvert + \lvert\gamma_C\rvert - \lvert\gamma_A\rvert}2.
\end{align}
We now consider an arbitrary configuration $\{g_x\}$ satisfying the boundary conditions \cref{eq:bc}.
We can then lower bound the energy as follows:
\begin{align}
  E[\{g_x\}]
&\geq \sum_{p \in P} \sum_{e=\braket{x,y} \in p} d(g_x,g_y)
\geq |P_{AB}| \, d(X,X^{-1}) + |P_{AC}| \, d(X,I) + |P_{BC}| \, d(X^{-1},I)
\nonumber \\
&=E_0^{(k)} \equiv \begin{cases}
\frac{k-1}2 (\lvert\gamma_A\rvert+\lvert\gamma_B\rvert+\lvert\gamma_C\rvert) & \text{$k$ odd}, \\
(\frac k2 - 1) (\lvert\gamma_A\rvert+\lvert\gamma_B\rvert) + \frac k2 \lvert\gamma_C\rvert & \text{$k$ even}.
\end{cases}
\label{eq:def E_0^(k)}
\end{align}
The first inequality is obtained by restricting to the edges that appear on paths in~$P$.
The second inequality follows by using the triangle inequality along each path and using the boundary conditions~\eqref{eq:bc}.
The final equality follows by \cref{eq:path count,eq:perm dists}.
Thus, $E_0^{(k)}$ is a lower bound on the energy of any configuration $\{g_x\}$.
When is this bound tight?
\begin{enumerate}
\item The first inequality is tight if and only if $g_x=g_y$ for every edge $e=\overline{xy}$ that is \emph{not} on a path in~$P$.
\item The second inequality is tight if and only if each path in~$P$ is a geodesic on~$S_k$ in the sense of \cref{eq:geodesic}.
\end{enumerate}
Thus, we can saturate the lower bound in the following way:
Let $\Gamma_A$, $\Gamma_B$, $\Gamma_C$ be minimal (and hence disjoint) min-cuts for $A$, $B$, and $C$, respectively, and define~$\Gamma'=V\setminus(\Gamma_A\cup\Gamma_B\cup\Gamma_C)$.
Define
\begin{align}\label{eq:min energy in ent wedges}
  g_x = \begin{cases}
    X, &x \in \Gamma_A, \\
    X^{-1}, &x \in \Gamma_B, \\
    \id, &x \in \Gamma_C,
  \end{cases}
\end{align}
and for~$x\in\Gamma'$, set $g_x$ to some permutation~$\tau$ that can simultaneously be on geodesics between $\id$, $X$, and $X^{-1}$ -- the same permutation for each connected component of $\Gamma'$.
Indeed, the first condition is satisfied since only edges in $\partial\Gamma_A$, $\partial\Gamma_B$, and $\partial\Gamma_C$ contribute, and those are covered by paths in~$P$ (\cref{eq:entropies}).
The second condition is satisfied by our choice of $g_x$ for $x\in\Gamma'$.
Thus, $E_0^{(k)}$ is the minimal energy of the spin model \cref{eq:spin model appendix,eq:bc}.

We claim that the configurations just described are the \emph{only} minimal energy configurations -- provided that the minimal cuts for~$A$, $B$, and~$C$ are unique.
This means that the degeneracy is given by \cref{eq:num tau}, the number of allowed $\tau$'s, to the power~$\numdomains$, the number of connected components of~$\Gamma'$.
As a consequence,
\begin{align}\label{eq:part func asymptotics}
  Z_k = a_k^{\numdomains} \times e^{-\log(D) E_0^{(k)}} \left( 1 + O(1/D) \right).
\end{align}
To prove the claim, start by defining the domains
\begin{align*}
  D_n(g) = \{ x\in V : d(g_x, g) \leq n \}.
\end{align*}
We claim that $D_n(X)$ is a min-cut for $A$ for each $n=0,\dots,n_A-1$, where $n_A=\left\lceil\frac k2\right\rceil-1$.
Indeed, $D_n(X) \cap V_\partial = A$ by the boundary conditions, so $D_n(X)$ is certainly a cut for~$A$.
Next, note that, by condition~1, any edge that leaves a domain of the form~$D_n(g)$ must necessarily be on a path in~$P$.
Now, any path in~$P_{AB}$ or~$P_{AC}$ necessarily starts in~$A \subseteq D_n(X)$ and, once it leaves $D_n(X)$, never returns to it.
This is because $d(X, g_x)$ is monotonically increasing along any geodesic path starting in~$A$.
Similarly, no path in~$P_{BC}$ can enter $D_n(X)$.
This is because if $x\in D_n(X)$ is on a path in~$P_{BC}$ then we would have
\begin{align*}
  k-1
&= d(X^{-1}, \id)
= d(X^{-1}, g_x) + d(g_x, \id) \\
&\geq \bigl(d(X^{-1}, X) - d(g_x, X)\bigr) + \bigl(d(X, \id) - d(X, g_x)\bigr) \\
&> d(X^{-1}, X) + d(X, \id) - 2n_A
= k-1
\end{align*}
where we first used that $g_x$ is on a geodesic from~$X^{-1}$ to~$\id$, next the triangle inequality (twice), and finally that $y\in D_n(x)$.
This is a contradiction.
Thus, each edge leaving $D_n(X)$ belongs to a distinct path in~$P_{AB}$ or~$P_{AC}$.
It follows that
\begin{align*}
  \partial(D_n(X)) \leq |P_{AB}| + |P_{AC}| = \lvert\gamma_A\rvert,
\end{align*}
which shows that $D_n(X)$ is a min-cut for~$A$.
Since we assumed that the min-cuts are unique, we find that $D_0(X) = D_{n_A-1}(X) = \Gamma_A$.
One similarly finds that $D_n(X^{-1}) = \Gamma_B$ for $n=0,\dots,n_B-1$ and $D_n(I) = \Gamma_C$ for $n=0,\dots,n_C-1$, where $n_B=\left\lceil\frac k2\right\rceil-1$ and $n_C=\left\lfloor\frac k2\right\rfloor$.
Thus, we find that any minimal energy configuration necessarily satisfies \cref{eq:min energy in ent wedges}.
What is more, for all bulk vertices in the complement of the three min-cuts, $x\in\Gamma'$, we have that
\begin{align}\label{eq:bulk perms ieq}
  d(\id, g_x) \geq \left\lfloor\frac k2\right\rfloor, \quad
  d(X, g_x) \geq \left\lceil\frac k2\right\rceil-1, \quad
  d(X, g_x) \geq \left\lceil\frac k2\right\rceil-1.
\end{align}
We still need to show that all three inequalities are saturated, as in \cref{eq:permtri eqv}, and that moreover $g_x$ is constant within each connected component of~$\Gamma'$.
For this, consider an arbitrary path in~$P_{AB}$, say.
The corresponding path of permutations~$g_x$ takes the form
It takes the form
\begin{align*}
  X \to \dots \to X \to g_1 \to \dots \to g_l \to X^{-1} \to \dots \to X^{-1},
\end{align*}
where the $X$ correspond to vertices in $\Gamma_A$, the $X^{-1}$ to vertices in $\Gamma_B$, and the $g_j$ to vertices in~$\Gamma'$.
Since the permutations along each path form a geodesic (by condition~2 above), we have that
\begin{align*}
  d(X,X^{-1})
&= d(X, g_j) + d(g_j, g_{j+1}) + d(g_{j+1}, X^{-1}) \\
&\geq \left\lceil\frac k2\right\rceil-1 + \left\lceil\frac k2\right\rceil-1 + d(g_j,g_{j+1})
= d(X,X^{-1}) + d(g_j,g_{j+1})
\end{align*}
where the inequality follows from \cref{eq:bulk perms ieq}.
Thus, all $g_x$ are the same along any path in $P_{AB}$ and equal to some permutation~$g$ that satisfies
\begin{align}\label{eq:along P_AB}
  d(X, g) = d(X^{-1}, g) = \left\lceil\frac k2\right\rceil-1.
\end{align}
Analogously, $g_x$ is also constant along any path in $P_{AC}$ and $P_{BC}$.
In the former case, it is equal to some permutation~$g$ that satisfies
\begin{align}\label{eq:along P_AC}
  d(X, g) = \left\lceil\frac k2\right\rceil-1, \quad d(\id, g) = \left\lfloor\frac k2\right\rfloor,
\end{align}
and in the latter case to one that satisfies
\begin{align}\label{eq:along P_BC}
  d(X^{-1}, g) = \left\lceil\frac k2\right\rceil-1, \quad d(\id, g) = \left\lfloor\frac k2\right\rfloor.
\end{align}
It follows that $\pi_x$ is constant on each connected component of $\Gamma'$.
Indeed, any edge~$\overline{xy}$ with endpoints in~$\Gamma'$ is either on a path in~$P$, so we can apply the preceding argument, or not, in which case we know that $\pi_x=\pi_y$ from condition~1 above.

To prove that all three inequalities in \cref{eq:bulk perms ieq} are saturated, it suffices to argue that any connected component~$\Gamma'$ meets paths in at least two out of $P_{AB}$, $P_{AC}$, and $P_{BC}$.
Indeed, \cref{eq:bulk perms ieq} is implied by any two out of the three equalities in \cref{eq:along P_AB,eq:along P_AC,eq:along P_BC} and the fact that $g_x$~is locally constant.
Assume for sake of contradiction that~$\Gamma'_0$ is a connected component of~$\Gamma$ that is only covered by paths in~$P_{AB}$, say.
This implies that $\partial\Gamma'_0$ can only connect to $\Gamma_A$ and $\Gamma_B$.
Clearly, each path in $P_{AB}$ can enter~$\Gamma'_0$ at most once from $\Gamma_A$ and then must leave $\Gamma'_0$ into $\Gamma_B$ (since we cannot have $X \to g \to X$ or $X^{-1} \to g \to X^{-1}$ on a geodesic).
This implies that $|\partial(\Gamma_A \cup \Gamma'_0)| = |\partial(\Gamma_A)|$, i.e., $\Gamma_A \cup \Gamma'_0$ is another min-cut for~$A$.
By uniqueness of the min-cut, $\Gamma'_0=\emptyset$, which is a contradiction.
We can similarly obtain a contradiction in the case that $\Gamma'_0$ meets only paths in~$P_{AC}$ or in~$P_{BC}$.
Thus, we find that every connected component of~$\Gamma'$ meets paths in at least two out of $P_{AB}$, $P_{AC}$, and $P_{BC}$.
As explained above, this means that it is filled with some fixed permutation~$g$ that satisfies \cref{eq:permtri,eq:permtri eqv}.
This concludes the proof.

\section{Asymptotic eigenvalue distribution of the partial transpose}\label{app:spec}
In this appendix, we compute the eigenvalue distribution of the partial transpose~$\rho_{AB}^{T_B}$ in the limit of large bond dimension $D\to\infty$, where $\rho_{ABC} \equiv \ket\Psi\bra\Psi$ and $\ket\Psi_{ABC}$ is the random tensor network state as defined in \cref{eq:rtns} of \cref{app:spin model}.
In particular, this gives a rigorous derivation of our asymptotic formula for the logarithmic negativity (\cref{eq:negativityasymptotics}).

To this end, we follow the method of moments (see, e.g.,~\cite[Sec.~2.1.2]{anderson2010introduction})and consider
\begin{align*}
    M_{AB} \equiv D^{\frac12\left( \lvert\gamma_A\rvert+\lvert\gamma_B\rvert+\lvert\gamma_C\rvert \right)} \, \rho_{AB}^{T_B}.
\end{align*}
We first note that
\begin{align}\label{eq:rank bound}
  \rk M_{AB} = \rk \rho_{AB}^{T_B} \leq \rk \rho_A \rk \rho_B \leq D^{\lvert\gamma_A\rvert + \lvert\gamma_B\rvert}.
\end{align}
The first inequality holds for any quantum state, while the second holds in any tensor network state.
Thus, only the first~$D^{\lvert\gamma_A\rvert + \lvert\gamma_B\rvert}$ many eigenvalues of $\rho_{AB}^{T_B}$ and $M_{AB}$ can be nonzero.
We therefore consider the truncated empirical eigenvalue distribution of~$M_{AB}$, defined as
\begin{align*}
  \mu_D \equiv \frac1{D^{\lvert\gamma_A\rvert + \lvert\gamma_B\rvert}} \sum_{i=1}^{D^{\lvert\gamma_A\rvert + \lvert\gamma_B\rvert}} \delta_{s_i},
\end{align*}
where $s_i$ denotes the $i$-th largest eigenvalue of~$M_{AB}$ and~$\delta_s$ the Dirac probability distribution.
Note that $\mu_D$ is a \emph{random} probability distribution.
Its $n$-th moment is the random variable
\begin{align*}
  \int d\mu_D(\lambda) \, \lambda^n
&= \frac1{D^{\lvert\gamma_A\rvert + \lvert\gamma_B\rvert}} \sum_{i=1}^{D^{\lvert\gamma_A\rvert + \lvert\gamma_B\rvert}} s_i^n
= \frac1{D^{\lvert\gamma_A\rvert + \lvert\gamma_B\rvert}} \tr (M_{AB})^n \\
&= \frac{D^{\frac n2( \lvert\gamma_A\rvert+\lvert\gamma_B\rvert+\lvert\gamma_C\rvert )}}{D^{\lvert\gamma_A\rvert + \lvert\gamma_B\rvert}} \tr (\rho_{AB}^{T_B})^n
= D^{( \frac n2 - 1 )( \lvert\gamma_A\rvert+\lvert\gamma_B\rvert ) + \frac n2 \lvert\gamma_C\rvert} \tr (\rho_{AB}^{T_B})^n.
\end{align*}
Let us compute its expectation value:
\begin{align*}
\overline{\int d\mu_D(\lambda) \, \lambda^n}
&= D^{( \frac n2 - 1 )( \lvert\gamma_A\rvert+\lvert\gamma_B\rvert ) + \frac n2 \lvert\gamma_C\rvert} \overline{\tr (\rho_{AB}^{T_B})^n} \\
&= D^{( \frac n2 - 1 )( \lvert\gamma_A\rvert+\lvert\gamma_B\rvert ) + \frac n2 \lvert\gamma_C\rvert} Z_n \left( 1 + O(1/D) \right) \\
&= D^{( \frac n2 - 1 )( \lvert\gamma_A\rvert+\lvert\gamma_B\rvert ) + \frac n2 \lvert\gamma_C\rvert} a_n^\numdomains D^{-E_0^{(n)}} \left( 1 + O(1/D) \right) \\
&=  \left( 1 + O(1/D) \right) \begin{cases}
D^{- \frac12 ( \lvert\gamma_A\rvert + \lvert\gamma_B\rvert - \lvert\gamma_C\rvert )} \left( n C_{(n-1)/2} \right)^\numdomains & \text{$n$ odd,} \\
C_{n/2}^\numdomains & \text{$n$ even.}
\end{cases}
\end{align*}
where we first used \cref{eq:spin model appendix} to compute $\overline{\tr (\rho_{AB}^{T_B})^k} = \braket{N'_k}$ and then \cref{eq:part func asymptotics,eq:def E_0^(k),eq:num tau} to evaluate the ground state energy and degeneracy of the spin model;
recall that~$\numdomains$ denotes the number of domains that remain when removing the min-cuts.
Accordingly we have,
\begin{align}\label{eq:moment mean lim}
  \lim_{D\to\infty} \overline{\int d\mu_D(\lambda) \, \lambda^n}
= \begin{cases}
1 & \text{$n$ odd and $\lvert\gamma_A\rvert + \lvert\gamma_B\rvert = \lvert\gamma_C\rvert$,} \\
0 & \text{$n$ odd and $\lvert\gamma_A\rvert + \lvert\gamma_B\rvert > \lvert\gamma_C\rvert$,} \\
C_{n/2}^\numdomains & \text{$n$ even,}
\end{cases}
\end{align}
using that $\lvert\gamma_A\rvert + \lvert\gamma_B\rvert \geq \lvert\gamma_C\rvert$, with equality only if $r=0$ (since by assumption all min-cuts are unique).
Next, we need to show that the variance of the $n$-th moment goes to zero.
For this we compute:
\begin{align*}
  \overline{\left(\int d\mu_D(\lambda) \, \lambda^n \right)^2}
&= D^{( n - 2 )( \lvert\gamma_A\rvert+\lvert\gamma_B\rvert ) + n \lvert\gamma_C\rvert} \, \overline{\tr (\rho_{AB}^{T_B})^n \ot (\rho_{AB}^{T_B})^n} \\
&= D^{( n - 2 )( \lvert\gamma_A\rvert+\lvert\gamma_B\rvert ) + n \lvert\gamma_C\rvert} \, \overline{\braket{\Psi_{ABC}^{\ot 2n} | P_A(Y) \ot P_B(Y^{-1}) \ot P_C(\id) | \Psi_{ABC}^{\ot 2n}}},
\end{align*}
where $Y = (1\,2\,\dots\,n) (n\!\!+\!\!1 \, n\!\!+\!\!2 \,\dots\,2n)$.
The right-hand side expectation can be computed by a similar spin model as in \cref{app:spin model,app:domains}, except that now the spin model takes values in~$S_{2n}$ and the boundary conditions are as follows:
\begin{align*}
  g_x = \begin{cases}
    Y, &x \in A, \\
    Y^{-1}, &x \in B, \\
    \id, &x \in C.
  \end{cases}
\end{align*}
By a similar analysis as in \cref{app:domains}, we find that the minimal energy configuration of this spin model are given by filling the min-cuts with the boundary conditions and the residual domains by permutations of the form $(\tau,\tau')$, see discussion at the end of \cref{app:permutations}.
Thus, the ground state energy is $2 E_0^{(n)}$ and the degeneracy is $a_n^{2\numdomains}$.
As a consequence,
\begin{align*}
  \overline{\left(\int d\mu_D(\lambda) \, \lambda^n \right)^2}
&= D^{( n - 2 )( \lvert\gamma_A\rvert+\lvert\gamma_B\rvert ) + n \lvert\gamma_C\rvert} \, a_n^{2\numdomains} D^{-2 E_0^{(n)}} \left( 1 + O(1/D) \right)
\end{align*}
and hence
\begin{align}\label{eq:moment var}
  \Var(\int d\mu_D(\lambda) \, \lambda^n)
&= D^{( n - 2 )( \lvert\gamma_A\rvert+\lvert\gamma_B\rvert ) + n \lvert\gamma_C\rvert} a_n^{2\numdomains} D^{-2 E_0^{(n)}} O(1/D)
= O(1/D)
\end{align}

We now compute the asymptotic eigenvalue distribution of the partial transpose using \cref{eq:moment mean lim,eq:moment var}.
Recall that the \emph{Wigner semicircle distribution} is the distribution~$\mu_W$ on~$[-2,2]$ with density~$d\mu_W/d\lambda = \frac1{2\pi} \sqrt{4-\lambda^2} \, \id_{\lvert\lambda\rvert\leq2}$.
Its odd moments vanish by symmetry, and its even moments are given by $\int d\mu_W(\lambda) \, \lambda^n = C_{n/2}$.
Let
\begin{align}\label{eq:lim measure}
  \mu_\infty = \begin{cases}
    \mu_W^{\ot\numdomains} & \text{$\numdomains>0$,} \\
    \frac12 \delta_1 + \frac12 \delta_{-1} & \text{$\numdomains=0$ and $\lvert\gamma_A\rvert + \lvert\gamma_B\rvert > \lvert\gamma_C\rvert$,} \\
    \delta_1 & \text{$\numdomains=0$ and $\lvert\gamma_A\rvert + \lvert\gamma_B\rvert = \lvert\gamma_C\rvert$,}
  \end{cases}
\end{align}
where $\mu_W^{\ot\numdomains}$ denotes the distribution of a product of $\numdomains$~independent random variables, each with distribution~$\mu_W$.
The distribution $\mu_\infty$ is supported in $[-2^\numdomains,2^\numdomains]$, and its moments are given by
\begin{align}\label{eq:moment semicircle power}
  \int d\mu_\infty(\lambda) \, \lambda^n = \begin{cases}
1 & \text{$n$ odd and $\lvert\gamma_A\rvert + \lvert\gamma_B\rvert = \lvert\gamma_C\rvert$,} \\
0 & \text{$n$ odd and $\lvert\gamma_A\rvert + \lvert\gamma_B\rvert > \lvert\gamma_C\rvert$,} \\
C_{n/2}^\numdomains & \text{$n$ even,}
\end{cases}
\end{align}
which agrees precisely with the limit in \cref{eq:moment mean lim}.
We claim that the measures $\mu_D$ converge weakly, in probability, to the distribution~$\mu_\infty$.
Thus we want to show that, for all $f\in C_b(\mathbb R)$,
\begin{align}\label{eq:weak lim claim}
  \int d\mu_D(\lambda) \, f(\lambda) \stackrel{\Pr}{\longrightarrow} \int d\mu_\infty(\lambda) \, f(\lambda)
\end{align}
as $D\to\infty$.
In fact, we will show this for all $f\in C(\mathbb R)$ with at most polynomial growth.
We follow the standard argument in~\cite[Sec.~2.1.2]{anderson2010introduction}.

We first show that it suffices to consider compactly supported test functions.
Indeed, by the Markov inequality, it holds for any~$\eps>0$ and~$B>0$ that
\begin{align*}
  \Pr(\int d\mu_D(\lambda) \, |\lambda|^k \id_{|\lambda| > B} > \eps)
\leq \frac1\eps \overline{\int d\mu_D(\lambda) \, |\lambda|^k \id_{|\lambda| > B}}
\leq \frac1{\eps B^k} \overline{\int d\mu_D(\lambda) \, \lambda^{2k}}
\end{align*}
and hence, using \cref{eq:moment mean lim},
\begin{align*}
  \limsup_{D\to\infty} \Pr(\int d\mu_D(\lambda) \, |\lambda|^k \id_{|\lambda| > B} > \eps)
\leq \limsup_{D\to\infty} \frac1{\eps B^k} \overline{\int d\mu_D(\lambda) \, \lambda^{2k}}
= \frac{C_k^r}{\eps B^k}
\leq \frac{4^{kr}}{\eps B^k}
= \frac1\eps \left( \frac{4^r} B \right)^k,
\end{align*}
where we used the bound $C_k \leq 4^k$ on the Catalan numbers.
If we choose $B\equiv4^\numdomains + 1$, then the left-hand side is nondecreasing with $k$, while the right-hand side converges to zero.
Thus:
\begin{align}\label{eq:limsup}
  \lim_{D\to\infty} \Pr(\int d\mu_D(\lambda) \, |\lambda|^k \id_{|\lambda| > B} > \eps) = 0 \qquad (\forall k).
\end{align}
Since the distribution~$\mu_\infty$ is supported in~$[-2^\numdomains,2^\numdomains]$, it follows that we only need to prove \cref{eq:weak lim claim} for continuous functions~$f$ that are compactly supported in~$[-B,B]$.
We will do so next.
By the Stone-Weierstrass approximation theorem, there exists, for any such~$f$ and~$\delta>0$, a polynomial~$p$ such that~$\max_{|\lambda| \leq B} |f(\lambda) - p(\lambda)| \leq \delta/8$.
Then,
\begin{align*}
&\quad \Pr(\lvert \int d\mu_D(\lambda) \, f(\lambda) - \int d\mu_\infty(\lambda) \, f(\lambda) \rvert > \delta) \\
&\leq \Pr(\lvert \int d\mu_D(\lambda) \, p(\lambda) \id_{|\lambda|\leq B} - \int d\mu_\infty(\lambda) \, p(\lambda) \rvert > \frac34\delta)
\leq P_1 + P_2 + P_3,
\end{align*}
where
\begin{align*}
P_1 &\equiv
\Pr(\lvert \int d\mu_D(\lambda) \, p(\lambda) \id_{|\lambda|> B} \rvert > \frac\delta4), \\
P_2 &\equiv \Pr(\lvert \int d\mu_D(\lambda) \, p(\lambda) - \overline{\int d\mu_D(\lambda) p(\lambda)} \rvert > \frac\delta4), \\
P_3 &\equiv \Pr(\lvert \overline{\int d\mu_D(\lambda) p(\lambda)} - \int d\mu_\infty(\lambda) \, p(\lambda) \rvert > \frac\delta4).
\end{align*}
Writing $P$ as a linear combination of a finite number of monomials, we see that~$P_1\to0$ by \cref{eq:limsup}, $P_2\to0$ by \cref{eq:moment var} and the Chebyshev inequality, and, finally, $P_3=0$ for~$D$ large enough by the convergence of moments in \cref{eq:moment mean lim} to \cref{eq:moment semicircle power}.
This concludes the proof of \cref{eq:weak lim claim}.

We remark that \cref{eq:weak lim claim} implies directly that for $D\to\infty$ the \emph{absolute} eigenvalue distribution, i.e., the eigenvalue distribution of $\lvert M_{AB} \rvert = D^{-\frac12( \lvert\gamma_A\rvert+\lvert\gamma_B\rvert+\lvert\gamma_C\rvert )} \lvert\rho_{AB}^{T_B}\rvert$ tends to $\mu_Q^{\ot r}$, where $\mu_Q$ is the \emph{quartercircle distribution} on $[0,2]$ with density $d\mu_Q/d\lambda = \frac1\pi \sqrt{4-\lambda^2} \id_{\lambda\in[0,2]}$.

\medskip

As an application, we calculate the asymptotic value of the logarithmic negativity:
\begin{align*}
  E_N(\rho_{AB})
&= \log \frac {\tr \lvert \rho_{AB}^{T_B} \rvert} {\tr \rho}
= -\frac{\log D}2 \left( \lvert\gamma_A\rvert+\lvert\gamma_B\rvert+\lvert\gamma_C\rvert \right) + \log \tr \lvert M_{AB} \rvert - \log \tr \rho \\
&= \frac{\log D}2 \left( \lvert\gamma_A\rvert+\lvert\gamma_B\rvert-\lvert\gamma_C\rvert \right) + \log \int d\mu_D(\lambda) \lvert\lambda\rvert - \log \tr \rho
\end{align*}
Using \cref{eq:weak lim claim,eq:lim measure}, we find that
\begin{align*}
  E_N(\rho_{AB}) - \frac{\log D}2 \left( \lvert\gamma_A\rvert+\lvert\gamma_B\rvert-\lvert\gamma_C\rvert \right)
\stackrel{\Pr}{\longrightarrow} \log \int d\mu_\infty(\lambda) \, \lvert\lambda\rvert
= \numdomains \log \frac8{3\pi},
\end{align*}
where we recall that $\numdomains$ denotes the number of residual domains.
This confirms \cref{eq:negativityasymptotics}.

If $B = \bar A$ then $C=\emptyset$ and $\numdomains=0$ (i.e., there is no~$\tau$-region) and hence there is no order-one correction in the large $D$ limit:
\begin{align*}
  E_N(\rho_{AB}) - \frac {\log D}2 \left( \lvert\gamma_A\rvert+\lvert\gamma_B\rvert-\lvert\gamma_C\rvert \right)
\stackrel{\Pr}{\longrightarrow} 0.
\end{align*}

\section{Non-maximally entangled link states}\label{app:link}
We now discuss an interesting variation of the random tensor network model, where the link states~$\ket{L_e}$ in \cref{eq:rtns} are no longer assumed to be maximally entangled.
See Eq.~(5.12) in \cite{hayden2016holographic} for a discussion of the R\'enyi entropies in this setting.
Here, the computation of the average
\begin{align*}
  \overline{N'_k} \equiv \overline{\braket{\Psi^{\ot k}|P_A(X) \ot P_B(X^{-1}) \ot P_C(\id) |\Psi^{\ot k}}}
\end{align*}
proceeds as before up until \cref{eq:same up to here}.
Now,
\begin{align*}
&\quad \tr\left[\ket{L_e}^{\ot k}\bra{L_e}^{\ot k} \bigl( P_{x,e}(g_x) \ot P_{y,e}(g_y) \bigr)\right] \\
&= \tr\left[\ket{L_e}^{\ot k}\bra{L_e}^{\ot k} \bigl( P_{x,e}(g_x g_y^{-1}) \ot I \bigr)\right] \\
&= \tr\left[\rho_e^{\ot k} P_{x,e}(g_x g_y^{-1}) \right]
\end{align*}
where we used that $\ket{L_e}^{\ot x}$ is permutation-invariant and we let $\rho_e$ denote the reduced density matrix of $\ket{L_e}$ on either subsystem.
Then taking the logarithm of the preceding equation leads to a spin model where the action per bond is computed by $J(g_x^{-1} g_y)$, where
\begin{align*}
    J(h)=-\log {\rm tr}\left(P(X) \rho_e^{\otimes k}\right).
\end{align*}
This confirms \cref{eq:action nonmaximal}.

Now we focus on the $\tau$ that are non-crossing pairings of \cref{app:permutations}.
For simplicity, let us assume that the states~$\rho_e$ are the same for each link $e=\overline{xy}$, and that moreover the entanglement spectrum is nontrivial, so that the R\'enyi entropies are strictly decreasing.
For the domain wall between $\mathbb{I}$ and $\tau$, each link contributes
\begin{align*}
    J(\tau)=\left\lfloor \tfrac k2\right\rfloor S_2\left(\rho_e\right)
\end{align*}
which is the same for all $\tau$. For the domain wall between $X=(1\,2\,\dots\,k)$, suppose it contains nontrivial disjoint cycles of length $k_1,k_2,...,k_c$ (with $k_i>1$). Then the contribution of each link is
\begin{align*}
    J\left(X^{-1}\tau\right) = \sum_{i=1}^c\left(k_i-1\right)S_{k_i}\left(\rho_e\right)
\end{align*}
Since $d(\tau,X)=\sum_i\left(k_i-1\right)=\left\lceil \frac k2\right\rceil -1$ (\cref{eq:permtri eqv}) and the R\'enyi entropy is monotonically non-increasing with $n$, we have $S_{k_i}(\rho_e) \geq S_{\left\lceil \frac k2\right\rceil}(\rho_e)$ and hence
\begin{align}\label{eq:first neighbor condition}
    J\left(X^{-1}\tau\right)
\geq \sum_{i=1}^c\left(k_i-1\right)S_{\left\lceil\frac k2\right\rceil}\left(\rho_e\right)
=\left(\left\lceil \tfrac k2\right\rceil -1\right)S_{\left\lceil\frac k2\right\rceil}\left(\rho_e\right).
\end{align}
Moreover, equality holds if and only if $X^{-1}\tau$ consists of a single $\left\lceil\frac k2\right\rceil$-cycle.
Similarly, for the domain wall between $X^{-1}$ and $\tau$, we find that
\begin{align}\label{eq:second neighbor condition}
    J\left(X\tau\right)
\geq \left(\left\lceil \tfrac k2\right\rceil -1\right)S_{\left\lceil\frac k2\right\rceil}\left(\rho_e\right),
\end{align}
with equality if and only if $X\tau$ consists of a single $\left\lceil\frac k2\right\rceil$-cycle.

We will now show that the \emph{nearest neighbor permutations} $\tau=(12)(34)\cdots$ and its cyclic permutations are the only non-crossing pairings that satisfy \cref{eq:first neighbor condition,eq:second neighbor condition} with equality, that is, are such that $X\tau$ and $X^{-1}\tau$ consist of a single $\lceil\frac k2\rceil$-cycle.
These $\tau$'s are then necessarily the only ones allowed in dominant configurations of the spin model.
It is easy to see that there are two such permutations if $k$ is even, and $k$ many such permutations if $k$ is odd.
Thus, for non-maximally entangled linked states, the degeneracy of the spin model is much smaller than in the maximally entangled case, whose degeneracy is exponentially large in~$k$ (see \cref{eq:num tau}).

To prove this claim, we first consider the case that $k=2n$ is even.
Note that any non-crossing pairing $\tau$ exchanges the subset of even with the subset of odd numbers in~$\{1,\dots,k\}$; the same is true for~$X$.
Therefore, $X\tau$ preserves the even and the odd numbers.
Since~$X\tau$ is also non-crossing, it follows that for $X\tau$ to consist of a single $\frac k2$-cycle we must have $X\tau=(1\,3\,\cdots\,k\!\!-\!\!1)$ or $X\tau=(2\,4\,\cdots\,k)$.
Thus, $\tau$ is either $(1 \, 2)(3 \, 4)\cdots(k\!\!-\!\!1 \, k)$ or $(2 \, 3)(4 \, 5)\cdots(k \, 1)$.
Since then $X^{-1}\tau$ consists likewise of a single $\frac k2$-cycle, this proves our claim when~$k$ is even.

Now suppose that $k=2n+1$ is odd.
Let $\tau$ be a non-crossing pairing such that both~$X\tau$ and~$X^{-1}\tau$ consist of a single $\lceil\frac k2\rceil$-cycle.
Since $k$ is odd, $\tau$ necessarily has a fixed point.
By cyclically permuting $\tau$, we may assume without loss of generality that $\tau(k)=k$.
Then~$\tau$ exchanges the subset of even with the subset of odd numbers in $\{1,\dots,k-1\}$.
It follows that~$X\tau$ preserves~$\{2,4,\dots,k-1\}$, hence also~$\{1,3,\dots,k\}$.
Since~$X\tau$ is also non-crossing, it follows that for $X\tau$ to consist of a single $\lceil\frac n2\rceil$-cycle we must have $X\tau=(1\,3\,\cdots\,k)$.
Thus, $\tau = (1 \, 2)(3 \, 4)\cdots(k\!\!-\!\!2 \, k\!\!-\!\!1)$.
Then $X^{-1}\tau$ consists likewise of a single $n$-cycle, concluding the proof our claim when~$k$ is odd.

\bibliographystyle{JHEP}
\bibliography{refs}

\end{document}